\documentclass[preprint,aps,floats,nofootinbib,amssymb,superscriptaddress]{revtex4}
\usepackage[sort&compress]{natbib}
\usepackage{epsf,epsfig}
\usepackage{amsmath}
\usepackage{hyperref}
\usepackage{subfigure}
\usepackage{graphicx}
\usepackage{color}
\usepackage{mathrsfs}

\newcommand{\be}{\begin{equation}}
\newcommand{\ee}{\end{equation}}
\newcommand{\bea}{\begin{eqnarray}}
\newcommand{\eea}{\end{eqnarray}}
\newcommand{\nn}{\nonumber}

\newcommand{\ba}{\begin{array}}
\newcommand{\ea}{\end{array}}
\newcommand{\bi}{\begin{itemize}}
\newcommand{\ei}{\end{itemize}}

\newcommand{\fracp}[2]{ \left( \frac{#1}{#2} \right)}
\newcommand{\fracb}[2]{ \left( \frac{#1}{#2} \right)}

\newcommand{\lrp}[1]{ \left( #1 \right) }
\newcommand{\lrb}[1]{ \left( #1 \right) }

\newcommand{\intdk}{ \int \frac{d^{4-\epsilon}k}{(2\pi)^{4-\epsilon}} }
\newcommand{\mhsq}{ M_{H}^{2} }
\newcommand{\mhfr}{ M_{H}^{4} }
\newcommand{\lfactor}{ \fracp{1}{4 \pi v^{2}}^{2} }

\makeatletter
\begin{document}

%-----------------------------------
% Preprint numbers
%-----------------------------------
\preprint{ICCUB-13-221}
\preprint{UB-ECM-PF-13/92}

%-----------------------------------
% Title
%-----------------------------------
\title{\vspace*{1.in} Radiative corrections to $W_LW_L$ scattering \\
 in composite Higgs models}

%-----------------------------------
% Authors
%-----------------------------------
\author{Dom\`enec Espriu}\affiliation{Departament d'Estructura i Constituents de la Mat\`eria,
Institut de Ci\`encies del Cosmos (ICCUB), \\
Universitat de Barcelona, Mart\'i Franqu\`es 1, 08028 Barcelona, Spain}
\author{Federico Mescia}\affiliation{Departament d'Estructura i Constituents de la Mat\`eria,
Institut de Ci\`encies del Cosmos (ICCUB), \\
Universitat de Barcelona, Mart\'i Franqu\`es 1, 08028 Barcelona, Spain}
\author{Brian Yencho}\affiliation{Departament d'Estructura i Constituents de la Mat\`eria,
Institut de Ci\`encies del Cosmos (ICCUB), \\
Universitat de Barcelona, Mart\'i Franqu\`es 1, 08028 Barcelona, Spain}
\vspace*{2cm}

\thispagestyle{empty}

%-----------------------------------
% Abstract
%-----------------------------------
\begin{abstract}
The scattering of longitudinally polarized electroweak bosons is likely to play an important  role in 
the elucidation of the fundamental nature of the Electroweak Symmetry 
Breaking sector and in determining the Higgs interactions with this sector. In this paper, by making use 
of the Equivalence Theorem, we determine the renormalization properties of the
electroweak effective theory parameters in a model with generic 
Higgs couplings to the $W$ and $Z$ bosons.  When the couplings 
between the Higgs and the electroweak gauge bosons deviate from their Standard Model values, additional 
counterterms of $O(p^4)$ in the usual chiral counting are required. 
We also determine in the same approximation 
the full radiative corrections to the $W_LW_L \to Z_LZ_L$ process in this type of models. 
Assuming custodial invariance, all the related processes can be easily derived from this amplitude.
\end{abstract}

\maketitle

%%%%%%%%%%%%%%%%%%%%%%%%%%%%%%%%%%%%%%%%%%%%%%%%%%%%%%%%%
\section{Introduction}\label{sec:intro}
Much of the current theoretical work concerning the LHC implications for the 
Electroweak Symmetry Breaking sector (EWSBS) focus on the 
deviations of the Higgs boson couplings to the electroweak gauge sector
rather than the self-couplings of the gauge bosons themselves\footnote{
Anomalous four-gauge-boson couplings have not been measured yet in LHC experiments at
the moment of writing this paper}.
Yet, any deviations of the former from their
Standard Model (SM) values turn out to have implications for the latter; they are intimately 
intertwined at loop level and should be understood together,
as unitarity considerations demand.
We seek in the present paper to provide a consistent framework for future studies of both 
in the scattering of longitudinally polarized electroweak gauge bosons.

In a previous paper\cite{EY} we have already examined the implications of unitarity in
the scattering of longitudinally polarized electroweak gauge bosons when---in addition to the usual  
SM lagrangian with a light scalar 
state (the Higgs particle with
$M_H\simeq 125$ GeV \cite{atlas,cms})---one includes an EWSBS assumed to be strongly interacting. 
This sector can be described at energies, $M_H^2 < s
< (4 \pi v)^{2}$ by an Electroweak Chiral Effective lagrangian
(EChL)~\cite{ECHL}. 
In \cite{EY} we included a set of $O(p^4)$ operators to
describe the strongly interacting EWSBS but assumed that the couplings between
the Higgs and the electroweak gauge bosons were indistinguishable from
the values that they take in the SM. The main purpose of the present work is
to relax this hypothesis.

A general chiral lagrangian with a
nonlinear realization of the  $SU(2)_L \times SU(2)_R$ symmetry up to
$O(p^4)$ terms and including a light Higgs is 
\bea
\label{eq:1}
\mathcal{L} & = &  - \frac{1}{2} {\rm Tr} W_{\mu\nu} W^{\mu\nu} - \frac{1}{4} {\rm Tr} B_{\mu\nu} B^{\mu\nu} \\ \nn
& & + \frac{1}{2} \partial_{\mu} h \partial^{\mu} h - h v \lrp{\lambda v^{2}+\mu^{2}} - 
\frac{1}{2} h^{2} \lrp{\mu^{2}+3v^{2} \lambda} - d_{3} (\lambda v) h^{3} - d_{4} \frac{1}{4} \lambda h^{4} \\ \nn 
& & + \frac{v^{2}}{4} \lrp{1+2 a\fracp{h}{v}+ b \fracp{h}{v}^{2}} {\rm Tr}\, D_{\mu}U^{\dagger}D^{\mu}U \\ \nn
& & + \sum_{i=0}^{13} \mathcal{L}_{i} + \mathcal{L}_{\rm GF} + \mathcal{L}_{\rm FP}.
\eea
Here, the $U$ field contain the three Goldstone bosons associated to the breaking of the global group 
to the custodial sub-group $SU(2)_V$
\be
\label{eq:2}
U = \exp \lrp{i~\frac{w \cdot \tau}{v} },
\ee
the $w$ being the three Goldstone boson fields\footnote{We shall denote 
by $z$ the neutral Goldstone boson. $w^\pm =(w^1\mp w^2)/\sqrt{2}$.}. The matrix $U$ transforms as $U\to LUR^\dagger$
under the action of the global group $SU(2)_L \times SU(2)_R$. The covariant derivative is 
defined as
\be
\label{eq:3}
D_{\mu} U =  \partial_{\mu}U + \frac{1}{2} i g W_{\mu}^{i} \tau^{i} U - \frac{1}{2} i g' B_{\mu}^{i} U \tau^{3}.
\ee
The Higgs field $h$ is a gauge and $SU(2)_L \times SU(2)_R$ singlet. The vacuum expectation value $v\simeq 250$ GeV gives
the right dimensions to the exponent in $U$. The terms $ \mathcal{L}_{\rm GF}$ and $\mathcal{L}_{\rm FP}$ in 
Eq. (\ref{eq:1}) correspond to the gauge-fixing and Faddeev-Popov pieces respectively, whereas the term
\be
\label{eq:4}
\sum_{i=0}^{13} \mathcal{L}_{i} = \sum_{i=0}^{13} a_i \mathcal{O}_{i}
\ee
includes a complete set of $CP$-even, local, Lorentz and gauge invariant operators, four-dimensional 
operators $\mathcal{O}_i$ constructed with the help of the field $U$, covariant derivatives 
and the $SU(2)_L\times U(1)_Y$ field strengths $ W_{\mu\nu}$ and  $B_{\mu\nu}$. 
A complete list can be found in \cite{ECHL} and also in \cite{EY}. 
While we will still restrict ourselves to a small subset of all possible general couplings 
we study those that are experimentally accessible now or in the near future.

In  Eq. (\ref{eq:1}) we have included with respect to  \cite{EY}  two extra parameters $a$ and $b$ controlling 
the coupling of the Higgs to the gauge sector. Following conventions in~\cite{composite}, 
we have also introduced two additional parameters
 $d_{3}$, and $d_{4}$ that are commonly used in composite Higgs scenarios. They
 parametrize the three- 
and four-point interactions of the Higgs field in an effective way. Needless to say that in a composite Higgs scenario
such as the one we have in mind the Higgs potential need not be renormalizable and higher powers of 
the field $h$ could appear. There could be additional interaction terms with the electroweak gauge sector
of $O(h^3)$ or higher. None of this should affect the results below. 
 
The SM  case corresponds to $a=b=d_{3}=d_{4}=1$ in Eq. (\ref{eq:1}). Current LHC results indicate that $a$ and $b$ 
are not too far from these SM values\cite{bounds}, but at present deviations from these SM values 
cannot be excluded. In \cite{EY} we assumed 
that the extended EWSBS would manifest itself only through the appearance of non-zero values for the $a_i$ 
$O(p^4)$ coefficients but $a$ and $b$ (as well as $d_3$ and $d_4$) were assumed 
to be very close to 1. This is the most conservative hypothesis.
However, even if $a\simeq b \simeq 1$, if the EWSBS is such that $O(p^4)$ operators
are present unitarity violations reappear at large
energies in a way apparently similar to what happens in models that were copiously studied in the 
past \cite{nohiggs} in the context of a very heavy Higgs or Higgsless theories.  

In \cite{EY} we calculated the scattering amplitudes using the longitudinal components of the vector
bosons themselves as external states, rather than the corresponding Goldstone bosons\footnote{In \cite{EY} we 
treated the tree-level and the imaginary part of the one-loop exactly, but we actually had to resort to
the Equivalence Theorem for the real part of the one-loop correction in order to keep the calculation manageable.} 
as it is customarily done when one takes  advantage of the Equivalence Theorem \cite{ET}. The reason to do so is that
at the energies being now explored at the LHC, corrections to the Equivalence Theorem can be of some relevance\cite{esma}.

We enforced unitarity through the use of the Inverse Amplitude Method \cite{iam}. 
We found that, even when including a light SM Higgs boson of mass $M_{H} = 125$~GeV, the unitarity analysis 
predicts the appearance of dynamical resonances in much of the parameter space of the higher-order coefficients.  
Their masses extend from as low as $300$~GeV to nearly as high as the cutoff of the method 
of $4\pi v\simeq 3$~TeV, with rather narrow widths typically of  order 1 to 10~GeV. In the absence of these resonances 
virtually all parameter space of the anomalous couplings could be excluded.  However,
we also showed that the actual signal strength of these resonances, when 
compared with current Higgs search data, is such that they are not currently being probed in LHC Higgs search data.
Yet, if anomalous vector boson couplings exist, the resulting dynamical resonances they predict 
should definitely be observable with future LHC data.

The study in \cite{EY} therefore showed that there is a direct connection ---also when 
a light Higgs is present--- between 
anomalous four gauge boson couplings and the underlying structure of dynamical resonances 
in the scalar and vector channels.
This emphasizes the importance of measuring these couplings (currently not yet observed at the LHC) to elucidate
the fundamental nature of the EWSBS. These measurements have to go hand-in-hand with the search for the putative 
additional resonances, bearing in mind that their peak heights and widths bear little resemblance to the Higgs signal (in 
the scalar sector) or even to what is expected in previously studied strongly interacting theories (particularly
in the vector channel). The reason being     
that the unitarization of the scattering amplitudes with a light Higgs profoundly changes the
resonance structure with respect to the Higgs-less (or a very heavy Higgs) 
scenario in extended scenarios of EWSBS. 
The situation could be also more intriguing if the hypothesis of setting $a$ 
and $b$ to their SM values, namely $a=b=1$ is relaxed as unitarity violations are already apparent at tree-level. 

Before the phenomenological analysis however, the case $a\ne1$ and $b\ne1$ requires a complete new study of the 
radiative corrections, including
a detailed study of the divergences and counterterms in this new scenario. This is part of the 
present work. We will also present a complete calculation of the one-loop $W_LW_L \to Z_L Z_L$ scattering amplitude 
(and by extension, upon use
of custodial symmetry, of all four longitudinal electroweak gauge boson couplings). The one-loop calculation 
will be done by making use the Equivalence Theorem \cite{ET}, where the 
longitudinal components are replaced by the
corresponding Goldstone bosons. This approximation is enough to derive the counterterms relevant for the process
being discussed. The calculation is 
done in the non-linear realization, discussed above, as this is
the natural language in composite Higgs models. Note that although $S$-matrix elements are independent of the particular
parametrization, renormalization constants need not be.  

Finally we mention that when computing electroweak gauge boson scattering amplitudes by making use of the 
Equivalence Theorem approximation, particularly if the calculation is done in the
gauge where the Goldstone bosons are massless, 
some subtleties appearing in a complete calculation are not present. For instance, the results are automatically
custodially invariant as one is assuming $g=g^\prime=0$. Crossing symmetry 
is also easily implemented by the usual 
exchanges of the Mandelstam variables. Therefore it is particularly simple to reproduce all amplitudes from 
the $ww \to zz$ one and, accordingly, only higher dimensional operators that are manifestly custodially invariant are
needed when moving away from the SM. However, in a full calculation of the  $W_LW_L \to Z_L Z_L$ amplitude,
including $O(g,g^\prime)$ corrections, new non custodially invariant operators would be required as counterterms. 
Furthermore crossing symmetry (although obviously still holding) is harder to implement (see e.g. the discussion in \cite{EY}).
We emphasize once more that none of this affects the determination of the counterterms derived in this paper.

%%%%%%%%%%%%%%%%%%%%%%%%%%%%%%%%%%%%%%%%%%%%%%%%%%%%%%%%%%%%
\section{Lagrangian and counterterms}\label{sec:lagrangian}
The lagrangian in Eq. (\ref{eq:1}) will be our starting point. The parameters there have to be considered as 
renormalized quantities. 
We trade $\mu$ for $\mhsq$ using $\mhsq \equiv (\mu^{2} + 3v^{2} \lambda)$. We will use a renormalization scheme
where the relation $M_H^2= 2\lambda v^2$ that holds true at tree level remains true for renormalized quantities. 

Next we have to consider the counterterm lagrangian. This will be
\bea
\mathcal{\delta L} &=& - h v \lrp{ \delta{\mhsq}-2 v^{2} \delta{\lambda} - 2\lambda \delta{v^{2}}} - \frac{1}{2} \delta{\mhsq} 
h^{2}  - d_{3} \lrp{\delta{\lambda} v + \frac{1}{2} \lambda v \frac{\delta{v^{2}}}{v^{2}}} h^{3} - d_{4} \frac{1}{4} \delta{\lambda} h^{4}
\nn \\ \nn & &
+ \lrp{2 \lrp{\delta{a} - \frac{1}{2} a \frac{\delta{v^{2}}}{v^{2}} }\fracp{h}{v}+ \lrp{\delta{b}- b \frac{\delta{v^{2}}}{v^{2}} } 
\fracp{h}{v}^{2}} \lrb{ \frac{v^{2}}{4} {\rm Tr}\, D_{\mu}U^{\dagger}D^{\mu}U }
\\ \nn & &
+ \lrp{1+2 a\fracp{h}{v}+ b \fracp{h}{v}^{2}}
\lrb{
\frac{v^{2}}{4}  {\rm Tr}\, D_{\mu}U^{\dagger}D^{\mu}U }_{\delta{v^{2}}}
\\ & &
+ \delta{a_{4}} \lrp{ {\rm Tr}\, \lrb{V^{\mu} V^{\nu}} }^{2} + \delta{a_{5}} \lrp{ {\rm Tr}\, \lrb{V^{\mu} V_{\mu}} }^{2}.
\label{eq:1.3}
\eea  
We have included the possible higher-order terms from the two $O(p^4)$ operators that are relevant for $W_LW_L$ scattering 
in the custodial limit, namely $\mathcal{L}_{4}$ and $\mathcal{L}_{5}$ (see e.g. \cite{EY} for details). We omit the pieces
that are not relevant for $W_LW_L$ scattering. In the treatment of this paper non-custodial $O(p^4)$ operators are not
needed. 

The counterterm lagrangian needs some explanation. To begin with, we have not introduced 
counterterms for $d_3$ and $d_4$ as they affect mostly
the renormalization of the Higgs self-interactions of which there is no experimental information at present.
Their renormalization should not affect the counterterms
that interest us most, namely those directly related to $W_LW_L$ scattering, such as $\delta a_4$ and $\delta a_5$. 
Secondly, there are additional $\delta{v^{2}}$ counterterms coming from the third line of Eq.~\ref{eq:1.3}
that depend on the number of factors of $v$ in the different terms of the $U$ expansion.  For instance, terms like
\be
\label{eq:1.4}
\frac{1}{2} \partial_{\mu} z \partial^{\mu} z +  \partial_{\mu} w^{+} \partial^{\mu} w^{-}
\ee
will have no corresponding counterterm because they contain no factor of $v$. On the other hand, terms with more 
than two $w$ fields will result in counterterms. For example, consider one term contributing to the 
four-point interaction
\be
\label{eq:1.5}
\fracp{1}{3 v^{2}} z \partial_{\mu} z  \lrb{\partial^{\mu} w^{+} w^{-} + \partial^{\mu} w^{-} w^{+}} \longrightarrow - 
\fracp{\delta{v^{2}}}{v^{2}} \fracp{1}{3 v^{2}} z \partial_{\mu} z \lrb{\partial^{\mu} w^{+} w^{-} + \partial^{\mu} w^{-} w^{+}}
\ee
In addition there are wave function renormalization constants for the Higgs field, $Z_H$, and for the Goldstone boson 
fields, $Z_w$. Note that there is no mass term (and no corresponding counterterm) for the Goldstone bosons 
as we shall consistently work in
the 't Hooft-Landau gauge, where Goldstone bosons are strictly massless. 
The renormalization conditions we will employ are that (i) the tadpoles vanish at one loop, (ii) the mass parameters 
are the on-shell masses, (iii) and that the relation $\lambda = \mhsq / (2 v^{2})$ is now true of the renormalized 
quantities, rather than the bare ones. We also note that condition (ii) only ends up effecting the 
Higgs mass counterterm, as the Goldstone bosons will remain massless independent of any corrections 
to the two-point function.

As indicated in the introduction we shall make use of the Equivalence Theorem to determine the counterterms and
the $W_LW_L$ scattering amplitude rather than using the actual gauge degrees of freedom. As far as the 
counterterms are concerned, this procedure is good enough to
give the correct renormalization of the parameters $a$, $b$, $a_4$ and $a_5$ that parametrize the EWSBS and thus
the departures from the SM result. As for the finite pieces of the amplitude, the use of the Equivalence Theorem 
is just an 
approximation\footnote{In \cite{EY} we used the Equivalence Theorem in the 't Hooft-Landau gauge to compute
the one-loop real part of the amplitude for simplicity. It was seen there that in spite of this approximation
unitarity was approximately preserved.} that becomes better for $s\gg M_W$. A complete calculation 
using the gauge degrees of freedom
is just too complicated for the present purposes and it is available numerically only for the SM\cite{DH}.

%%%%%%%%%%%%%%%%%%%%%%%%%%%%%%%%%%%%%%%%%%%%%%%%%%%%%%%%%%%%%%%%%%%%%%%%%%%%%%%%%%%%%%%%%%%%%%
\section{Tree-level calculation of $w^+ w^- \to zz$}\label{sec:tree-level}
\begin{figure}[tb]
\centering
\subfigure[(a)]{\includegraphics[clip,width=0.30\textwidth]{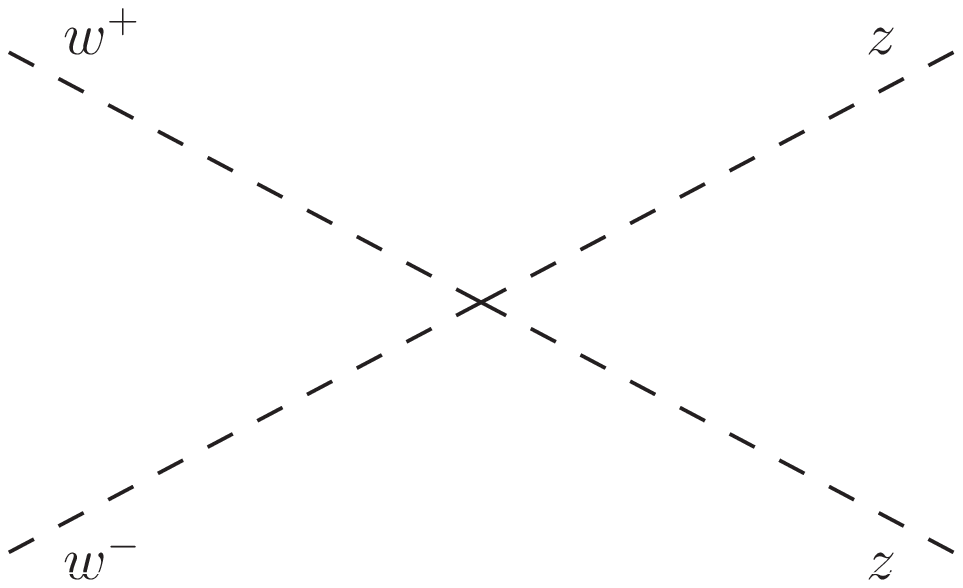}} \hspace{1cm}
\subfigure[(b)]{\includegraphics[clip,width=0.30\textwidth]{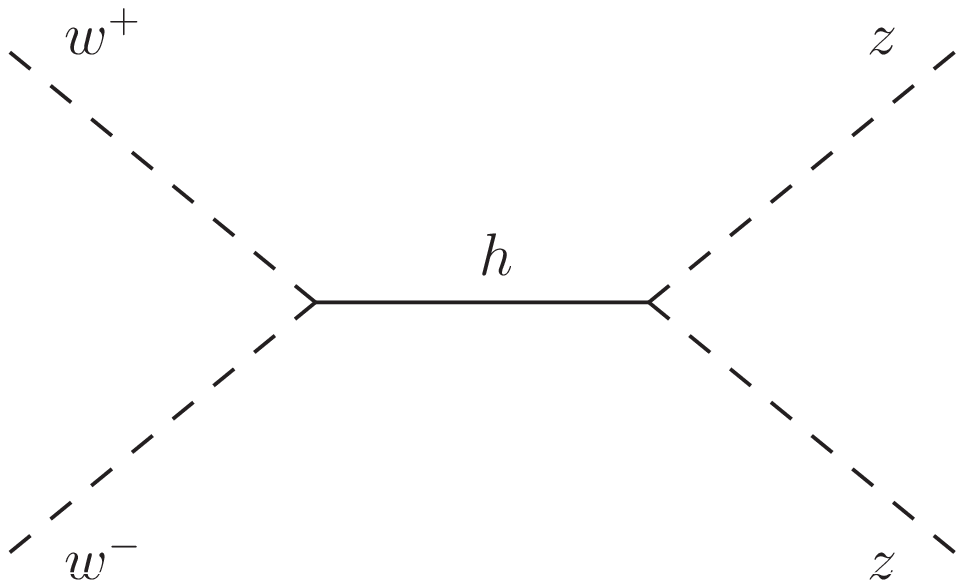}}
\caption{Tree-level diagrams contributing to the amplitude $w^+ w^- \to zz$, $i \mathcal{M}_{\rm tree}$}
\label{fig:diagrams_tree}
\end{figure}

The tree level calculation is fairly straightforward and comes from the sum of the two diagrams as in the usual 
linear realization case, albeit with different couplings: the $wwzz$ 4-pt diagram, 
and the $s$-channel Higgs exchange diagram. These diagrams are shown in Fig.~\ref{fig:diagrams_tree}. 
Their respective contributions are
\be
\label{eq:2.1}
i \mathcal{M}_{\rm tree}^{(a)} = i \fracp{s}{v^{2}}, {\text{ and }}
i \mathcal{M}_{\rm tree}^{(b)} = - i \fracp{a^{2} s}{v^{2}} \fracp{s}{s-\mhsq}.
\ee
Combined they give
\be
\label{eq:2.2}
i \mathcal{M}_{\rm tree} = - i \fracp{s}{v^{2}} \fracp{(a^{2}-1)s+\mhsq}{s-\mhsq},
\ee
which obviously reduces to the same value as the linear case for the SM ($a=1$). 
Note that in the following the assumption that $p_{i}^{2}=0$ 
is already made when presenting the amplitude.
This expression shows clearly
the $\sim s^2$ growth of the tree-level amplitude as $s \gg M_H^2$ if $a\neq 1$ signaling the
breakdown of unitarity already at tree-level when one moves away from the SM.

%%%%%%%%%%%%%%%%%%%%%%%%%%%%%%%%%%%%%%%%%%%%%%%%%%%%%%%%%%%%%%%%%%%%%%%%%%%%%%%%
\section{One-loop level calculation of $w^+ w^- \to zz$}\label{sec:loop-level} 
In the following, the classification of diagrams roughly follows the conventions given in ref.~\cite{DW}, but of course the 
calculation is completely different as the non-linear realization
is used in the present paper and additional topologies of the diagrams do appear.  
Single diagram includes contributions from internal $h$, $w^{\pm}$, 
and $z$ loops. We labelled by $(a)$  the subdiagrams for the $h$ loops and  
by $(b)$ the combined ones for $w^{\pm}$ and $z$ loops. 

Here, we will present the radiative corrections to the process grouped in several classes. There are the 
Higgs self-energy corrections to the diagram in Fig.\ref{fig:diagrams_two} and the vertex corrections in Fig.\ref{fig:diagrams_three}. Then we have some irreducible diagrams
that following~\cite{DW} we classify as bubbles (in Fig.\ref{fig:diagrams_bubbles}), triangles (in Fig.\ref{fig:diagrams_triangle}) and boxes (in Fig.\ref{fig:diagrams_box}). 
In addition we have two new type  of diagrams 
that appear only in the non-linear realization and thus have no counterpart in ref.~\cite{DW}. We have called them
five-field (in Fig.\ref{fig:diagrams_five}) and six-field (in Fig.\ref{fig:diagrams_six})  diagrams, respectively.

\subsection{Higgs self-energy corrections}
\label{sec:two_point}

\begin{figure}[tbh]
\centering
\subfigure[(a)]   {\includegraphics[clip,width=0.30\textwidth]{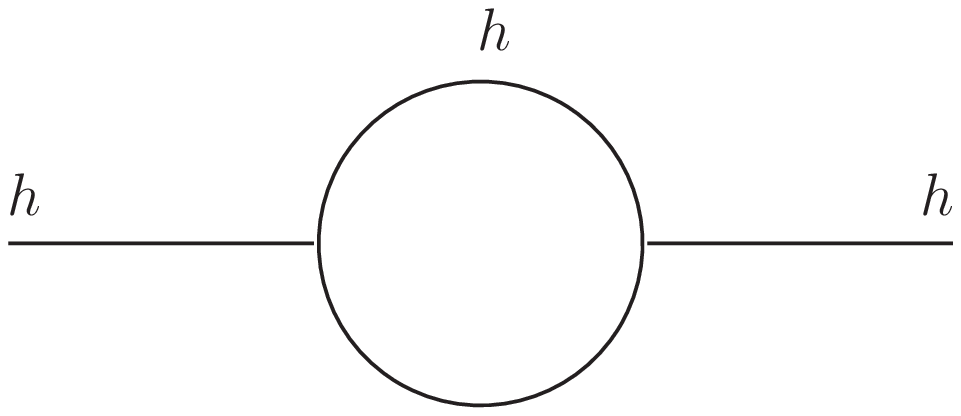}} \hspace{1cm}
\subfigure[(b)]{\includegraphics[clip,width=0.30\textwidth]{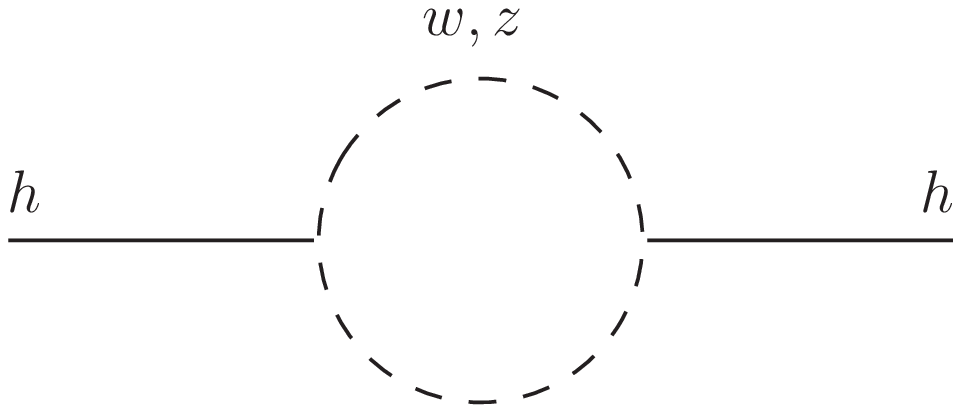}} \\
\subfigure[(c)]    {\includegraphics[clip,width=0.30\textwidth]{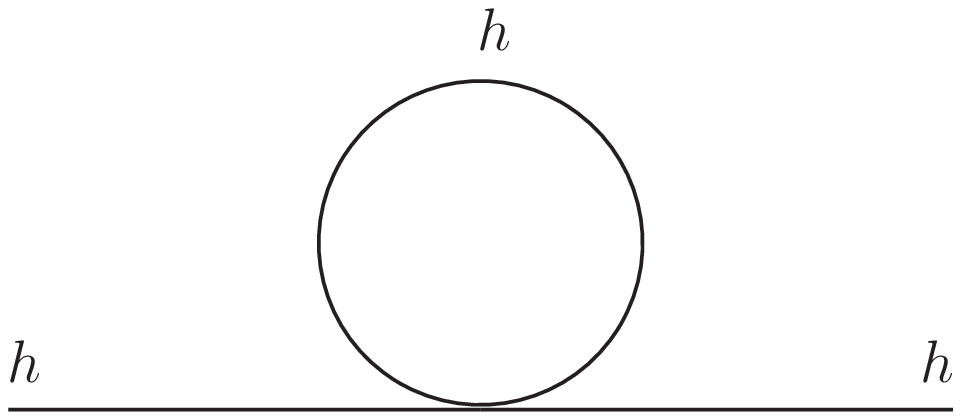}} \hspace{1cm}
\subfigure[(d)]  {\includegraphics[clip,width=0.30\textwidth]{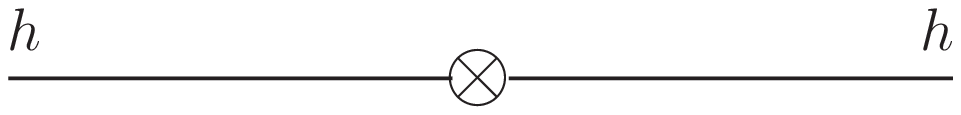}}
\caption{Radiative corrections to the Higgs two-point function}
\label{fig:diagrams_two}
\end{figure}
The two-point diagrams given in~\cite{DW} correspond to $-i \Pi(s)$ and are plotted in Fig.~\ref{fig:diagrams_two}. Their contribution to the tree-level diagram  $w^{+}w^{-} \to h \to zz$ can be parametrized as
\be
\label{eq:3.1}
i \mathcal{M}_{2-pt} = \fracp{a}{v}^{2} \frac{(s)^{2}}{(s-\mhsq)^{2}} \left[ -i \Pi(s) \right].
\ee
and for $d_{3}=d_{4}=1$ we have 
\bea
\label{eq:3.2}
i \mathcal{M}_{2-pt}
&=& i \lfactor \fracp{3~a^{2} \mhfr}{2} \frac{s^{2}}{\lrp{s-\mhsq}^{2}} \times \\ \nn 
& &
\Bigg(
\frac{A_{0}(\mhsq)}{\mhsq} + 3 B_{0}(s,\mhsq,\mhsq)
+ a^{2} \frac{s^{2}}{\mhfr} B_{0}(s,0,0)
\Bigg) \\ \nn 
& & 
- i \delta{\mhsq} \fracp{a}{v}^{2} \frac{s^{2}}{(s-\mhsq)^{2}}. 
\eea
The scalar functions $A_0$ and $B_0$ are described in the appendix and both are ultraviolet divergent. 
Note that the calculation 
includes the counterterm for
$\delta M_H^2$ (last line).

\subsection{$hw^+ w^-$ and $hzz$ vertex corrections}

\begin{figure}[tb]
\centering
\subfigure[(a)]   {\includegraphics[clip,width=0.30\textwidth]{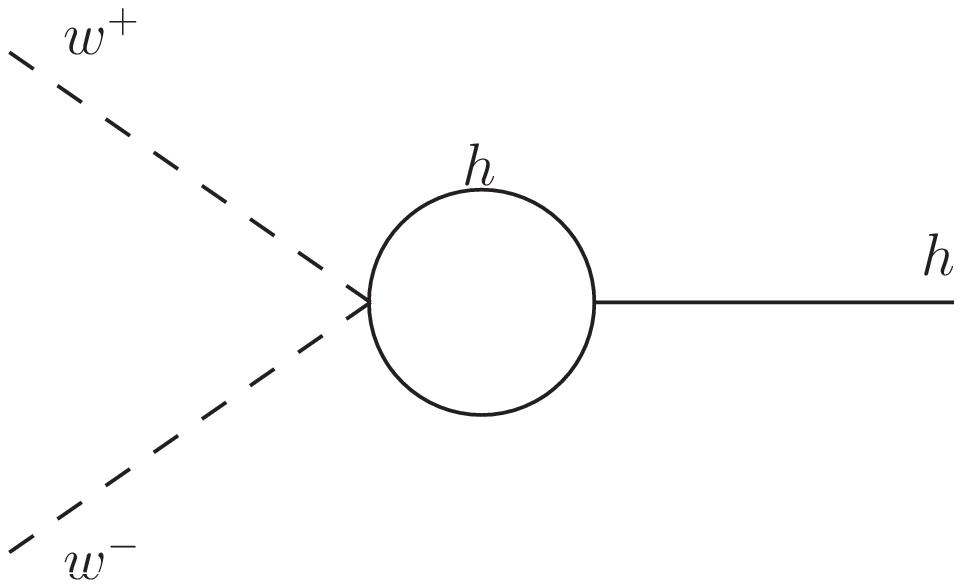}} \hspace{1cm}
\subfigure[(b)]{\includegraphics[clip,width=0.30\textwidth]{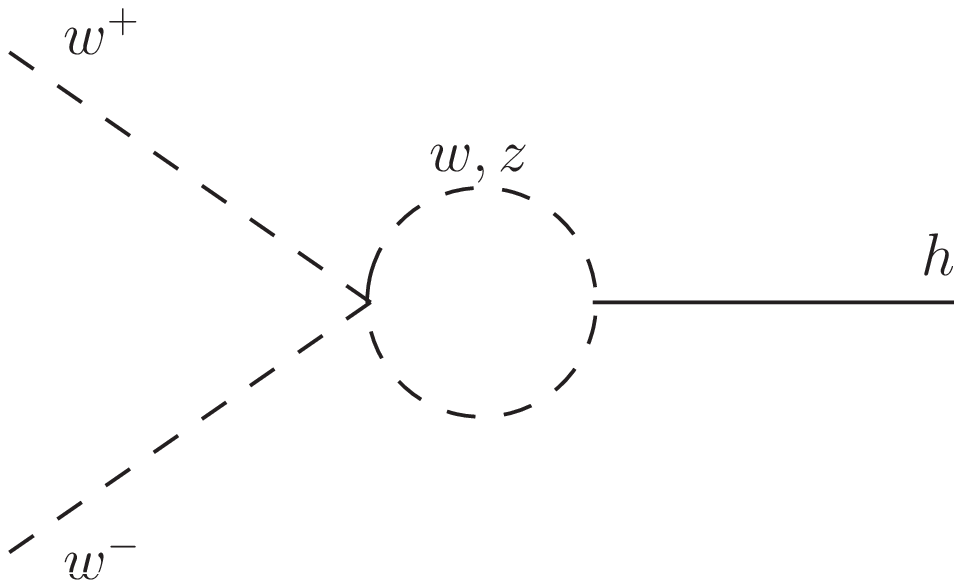}} \\
\subfigure[(c)]    {\includegraphics[clip,width=0.30\textwidth]{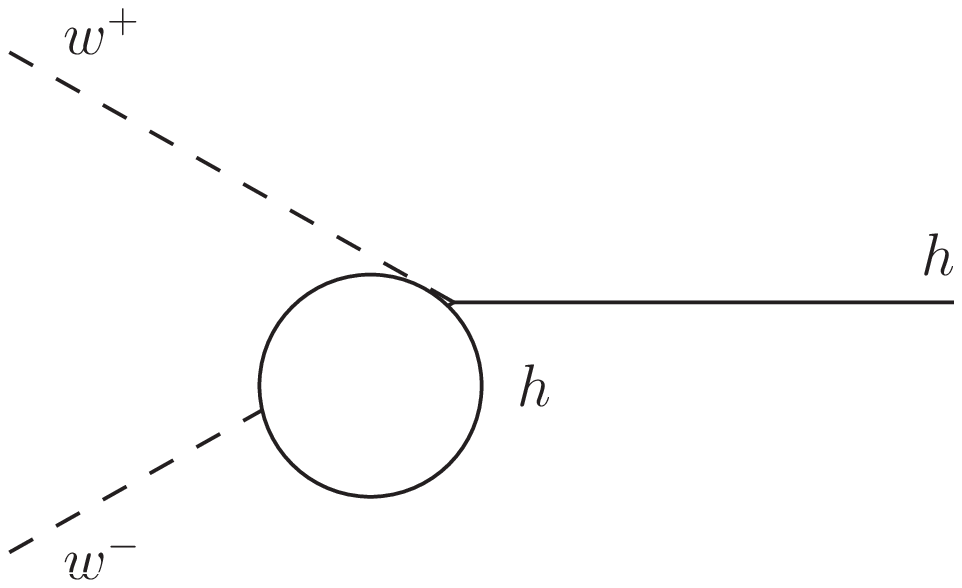}} \hspace{1cm}
\subfigure[(d)]    {\includegraphics[clip,width=0.30\textwidth]{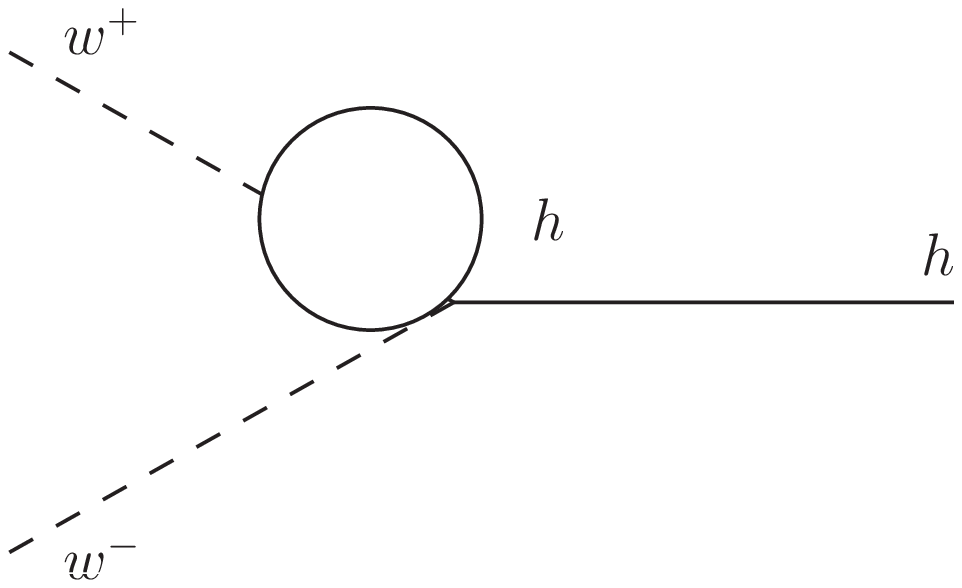}} \\
\subfigure[(e)]    {\includegraphics[clip,width=0.30\textwidth]{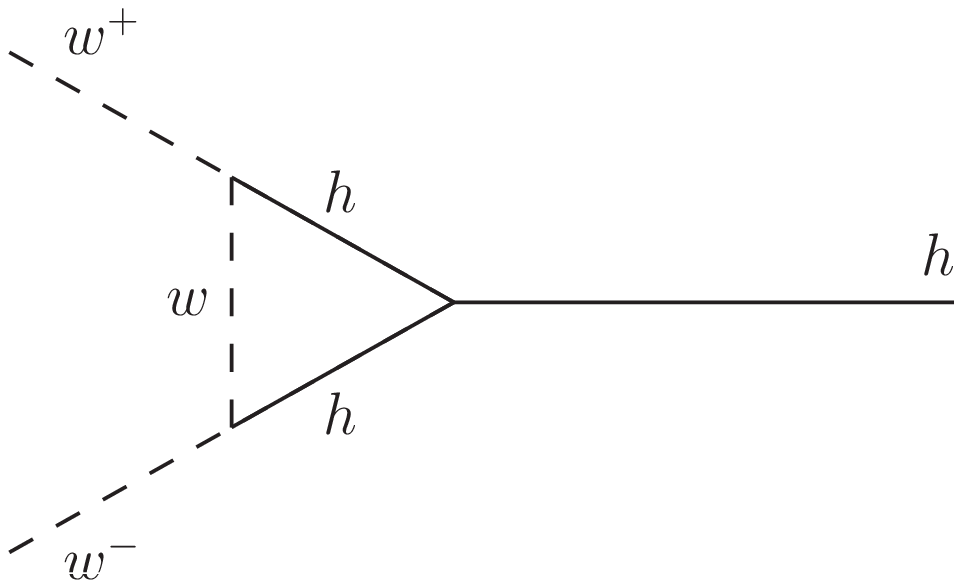}} \hspace{1cm}
\subfigure[(f)]    {\includegraphics[clip,width=0.30\textwidth]{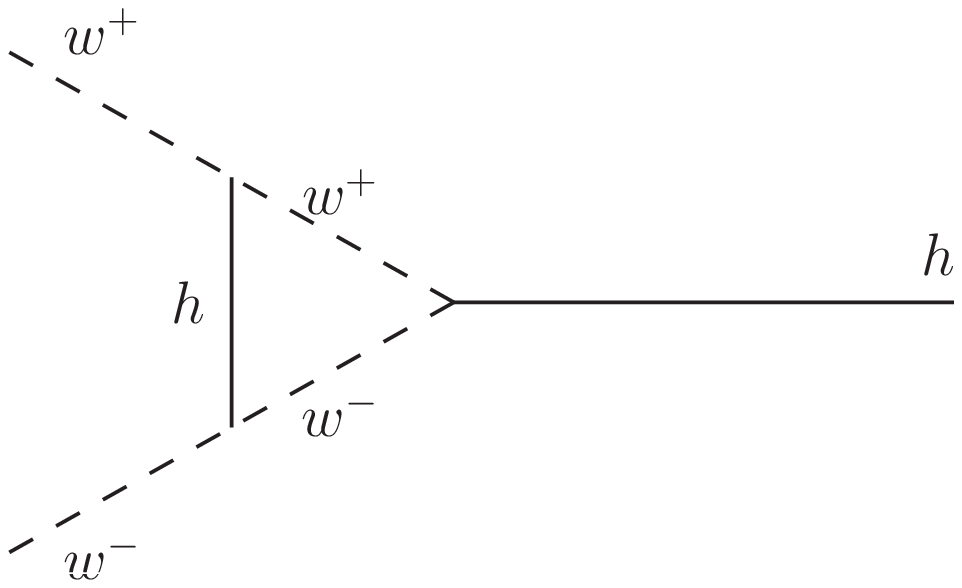}} \\
\subfigure[(g)]  {\includegraphics[clip,width=0.30\textwidth]{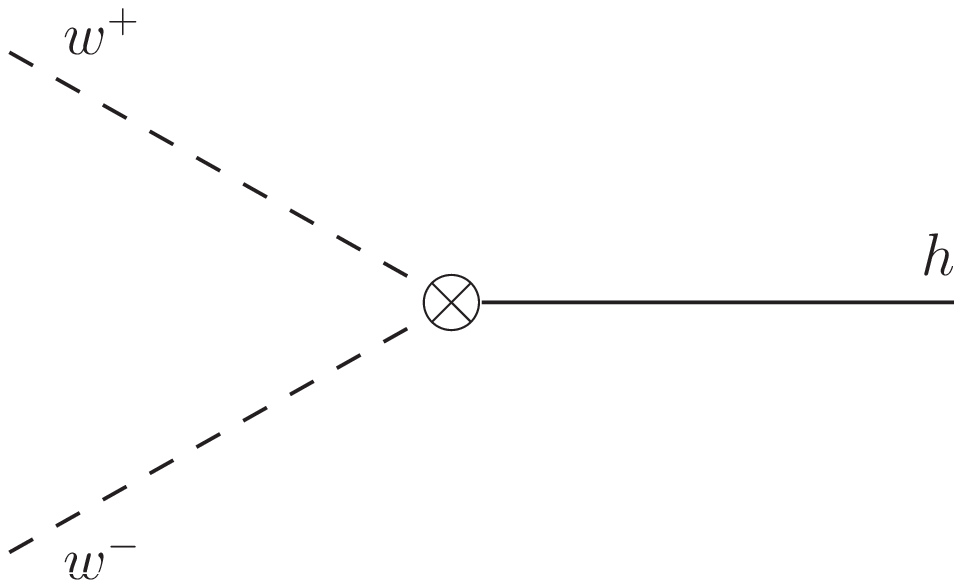}}
\caption{Three point vertex correction for the $h w^+w^-$ vertex. A slightly different set of diagrams for the vertex $hzz$ but the result is actually the same.}
\label{fig:diagrams_three}
\end{figure}

The three-point diagrams given in~\cite{DW} correspond to the $hww$/$hzz$ vertex correction $i \Gamma_{3}$, which 
is also related to the one-loop corrections to the Higgs decay width to $ww$/$zz$.  The total correction is the 
same for both the $hww$ and $hzz$ vertices, although the actual set of diagrams is slightly different for each 
in the non-linear representation, as there is a 4-$w$ coupling but no 4-$z$ coupling.   
We draw in Fig.\ref{fig:diagrams_three} diagrams for the case of the $hw^+ w^-$ vertex.
 Replacing appropriately $w$'s by $z$'s lines, we get the diagrams for the $hzz$ vertex. 
In this case, however,  we only have $z$ internal loops in Fig.\ref{fig:diagrams_three}(b).  
The rest of the diagrams are the same, however,  the total correction 
can be given as twice the correction to any one vertex to give
\be
\label{eq:3.3}
i \mathcal{M}_{3-pt} = \fracp{2a}{v} \frac{s}{(s-\mhsq)} \left[ i \Gamma_{3} \right]
\ee
We then have (for $d_{3}=d_{4}=1$) the total contribution 
\bea
\label{eq:3.4}
\!\!\!\!\!\!\!\!i \mathcal{M}_{3-pt} &=& 
-i \lfactor \lrp{a \mhfr} \fracp{s}{s-\mhsq} \times
\Bigg( 
- a (a^{2}-b) \fracp{s}{\mhsq} \Bigg.\\\nn
&& + 2 a \fracp{b s - 3 a \mhsq}{\mhsq} \frac{A_{0}(\mhsq)}{\mhsq}
+ 3 \fracp{(a^{2}-b)s + 2 a^{2} \mhsq}{\mhsq} B_{0}(s,\mhsq,\mhsq)
\\ \nn & &
+ a \fracp{s}{\mhsq} \fracp{(2+a^{2})s - 2 a^{2} \mhsq}{\mhsq} B_{0}(s,0,0)
\\ \nn & &
+  2 \lrp{a^{3} s} C_{0}(0,0,s,0,\mhsq,0)
-  6 \lrp{a^{2} \mhsq} C_{0}(0,0,s,\mhsq,0,\mhsq)
\Bigg)
\\ \nn & &
 -i (\delta{a}) \fracp{2 a s}{v^{2}} \frac{s}{(s-\mhsq)}  + i \fracp{\delta{v^{2}}}{v^{2}} \fracp{a}{v}^{2} \frac{s^{2}}{(s-\mhsq)}
\eea
Note the inclusion of the counterterms for the parameter $a$ (describing departures from the SM $hww$ and $hzz$ couplings
in the non-linear realization) and for the scale $v^2$. The (finite) scalar function $C_0$ is described in the appendix.

\subsection{Bubble diagrams}

\begin{figure}[tbh]
\centering
\subfigure[(a)]   {\includegraphics[clip,width=0.30\textwidth]{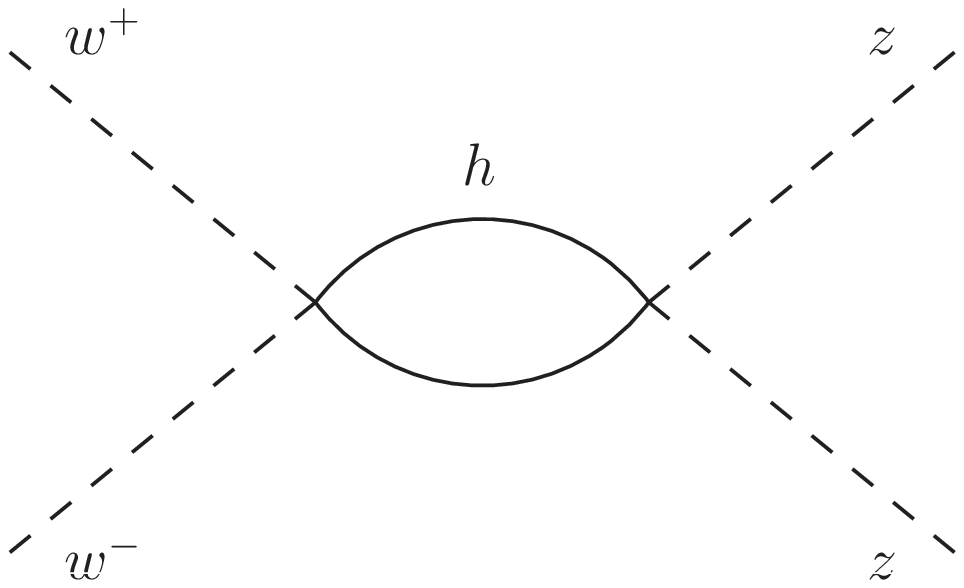}} \hspace{1cm}
\subfigure[(b)]   {\includegraphics[clip,width=0.30\textwidth]{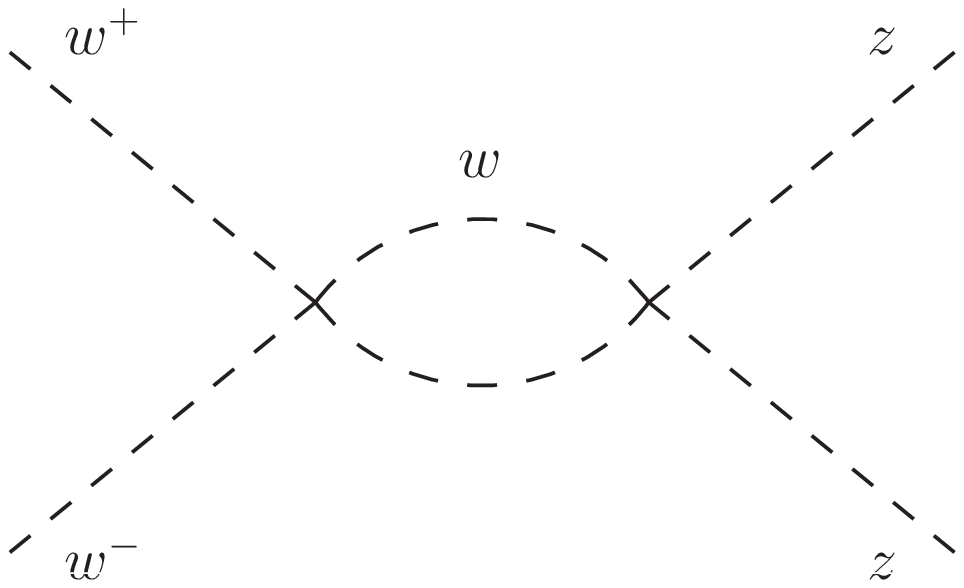}} \\
\subfigure[(c)]    {\includegraphics[clip,width=0.30\textwidth]{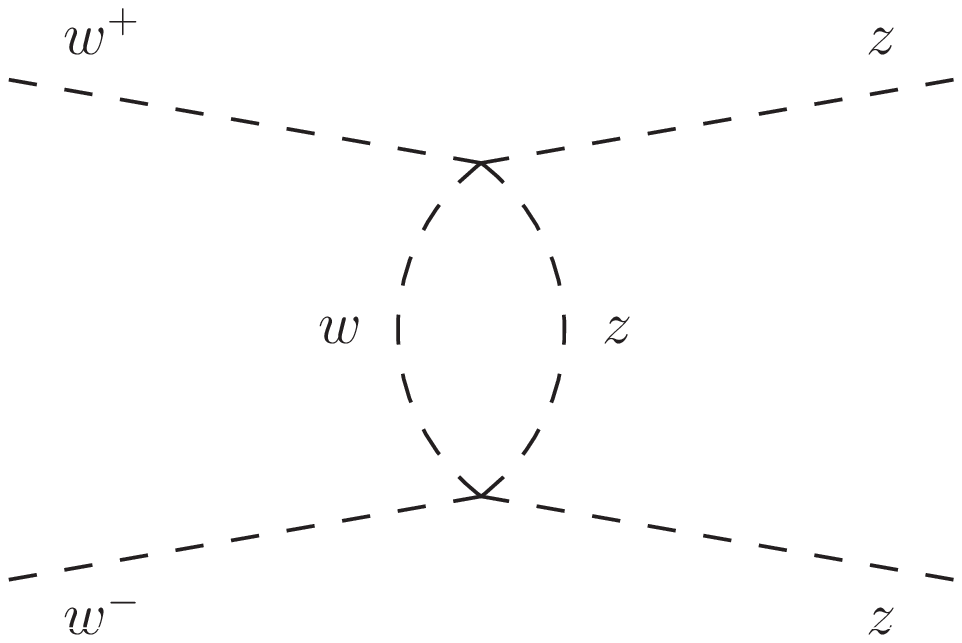}} \hspace{1cm}
\subfigure[(d)]    {\includegraphics[clip,width=0.30\textwidth]{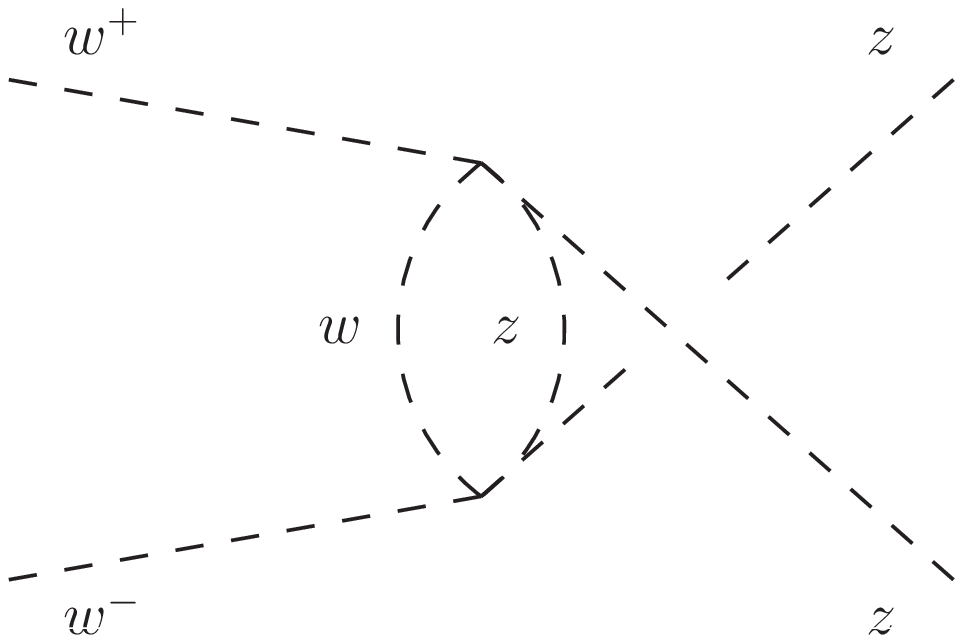}} \\
\subfigure[(e)]  {\includegraphics[clip,width=0.30\textwidth]{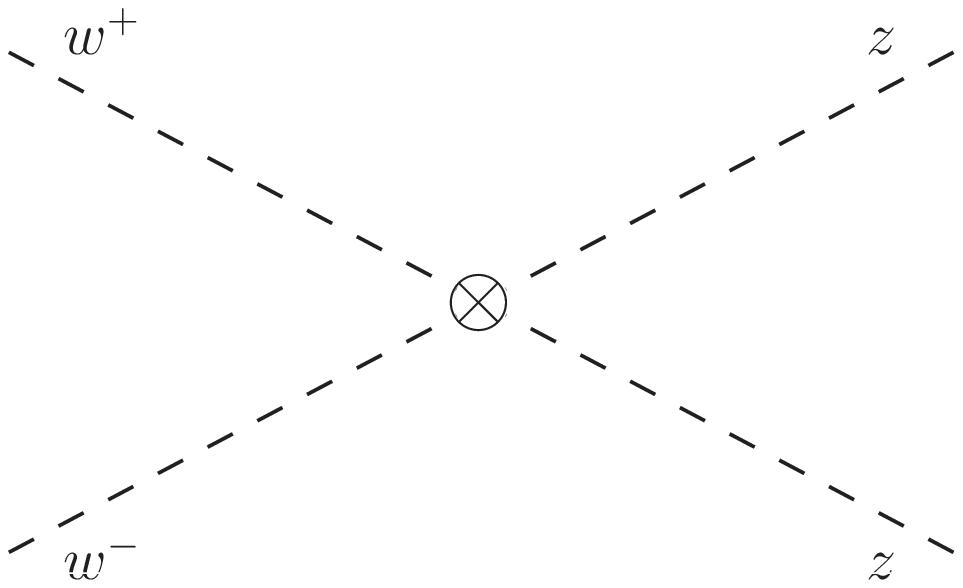}}
\caption{Bubble diagrams, $i \mathcal{M}_{bubbles}$. Note that we have included the four-point counterterms $\delta a_4 $ and
$\delta a_5$ here, but this is simply a choice.}
\label{fig:diagrams_bubbles}
\end{figure}

The bubble diagrams are given in Fig.~\ref{fig:diagrams_bubbles} and their contributions for $d_{3}=d_{4}=1$ sum up to
\bea
\label{eq:3.5}
i \mathcal{M}_{bubbles} &=& 
i \lfactor \fracp{\mhfr}{2} \times 
\\ \nn & &
\Bigg( 
  \fracp{-2 s^{2} + t^{2} + u^{2}}{9 \mhfr}
+ \fracp{s^{2}}{\mhfr} \Big(b^{2}  B_{0}(s,\mhsq,\mhsq) + B_{0}(s,0,0)\Big)
\\ \nn & &
+ \fracp{t (t-u)}{3 \mhfr} B_{0}(t,0,0)
+ \fracp{u (u-t)}{3 \mhfr} B_{0}(u,0,0)
\Bigg)
\\ \nn & &
 -i \fracp{s}{v^{2}} \fracp{\delta_{v^{2}}}{v^{2}}
+ i \fracp{4}{v^{4}} \lrb{ (\delta{a_{4}}) \lrp{t^{2}+u^{2}}+2(\delta{a_{5}}) \lrp{s^{2}}}
\eea
Note the inclusion here of the  counterterms for the $O(p^4)$ coefficients $a_4$ and $a_5$

\subsection{Triangle diagrams}

\begin{figure}[tbh]
\centering
\subfigure[(a)]{\includegraphics[clip,width=0.30\textwidth]{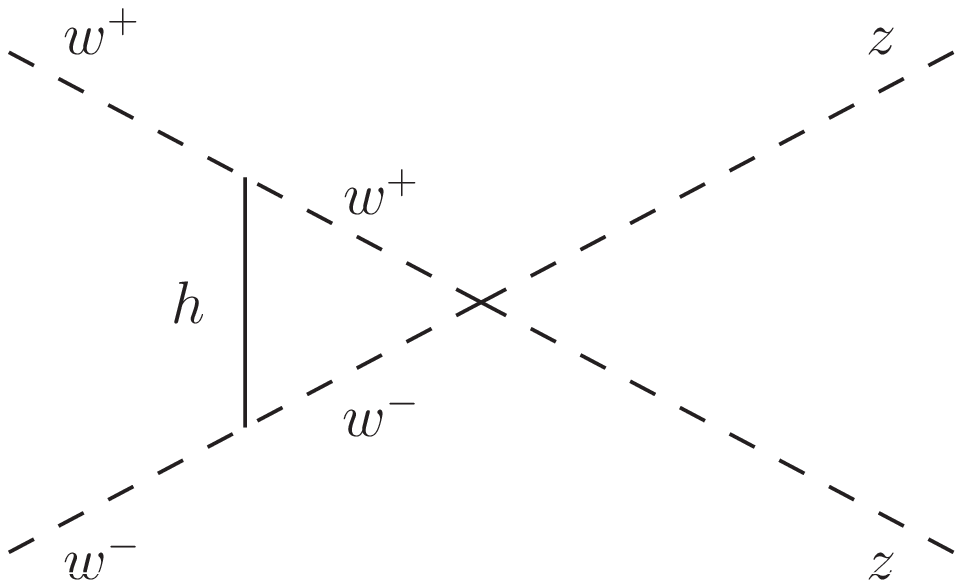}} \hspace{1cm}
\subfigure[(b)]{\includegraphics[clip,width=0.30\textwidth]{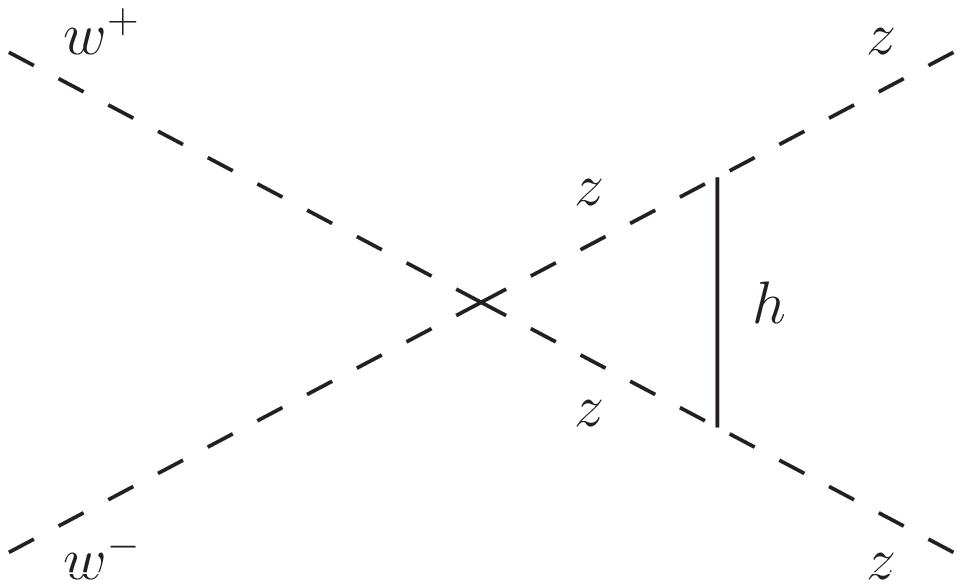}} \\
\subfigure[(c)]{\includegraphics[clip,width=0.30\textwidth]{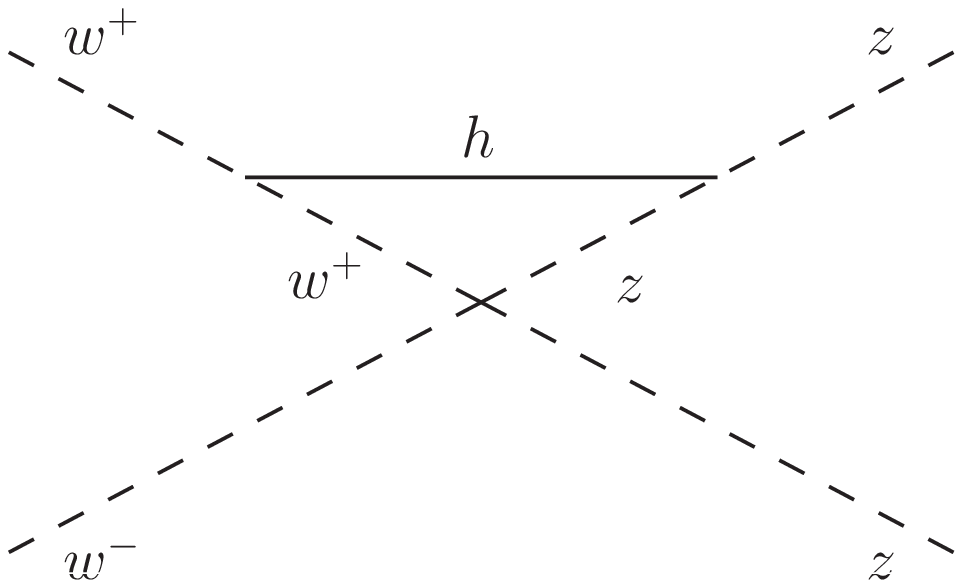}} \hspace{1cm}
\subfigure[(d)]{\includegraphics[clip,width=0.30\textwidth]{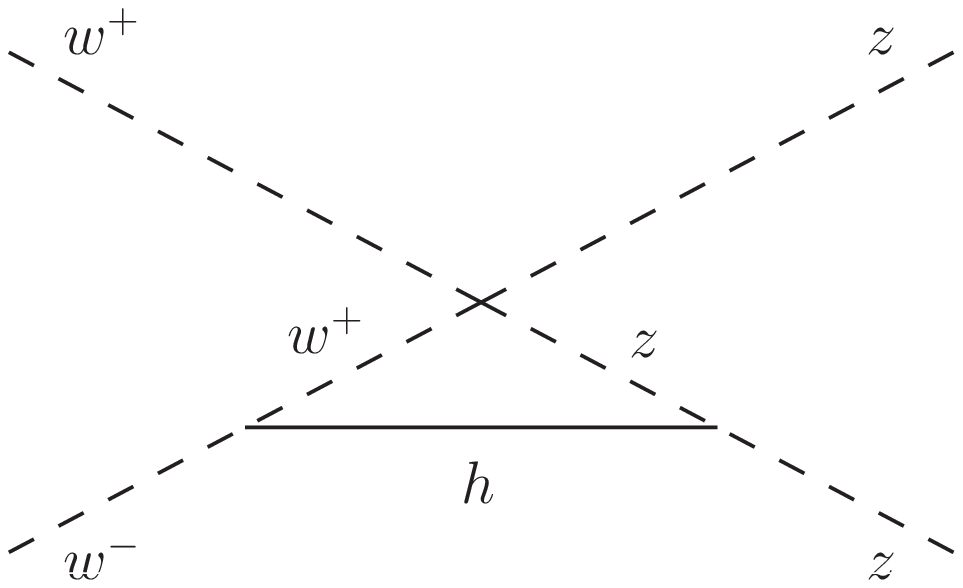}} \\
\subfigure[(e)]{\includegraphics[clip,width=0.30\textwidth]{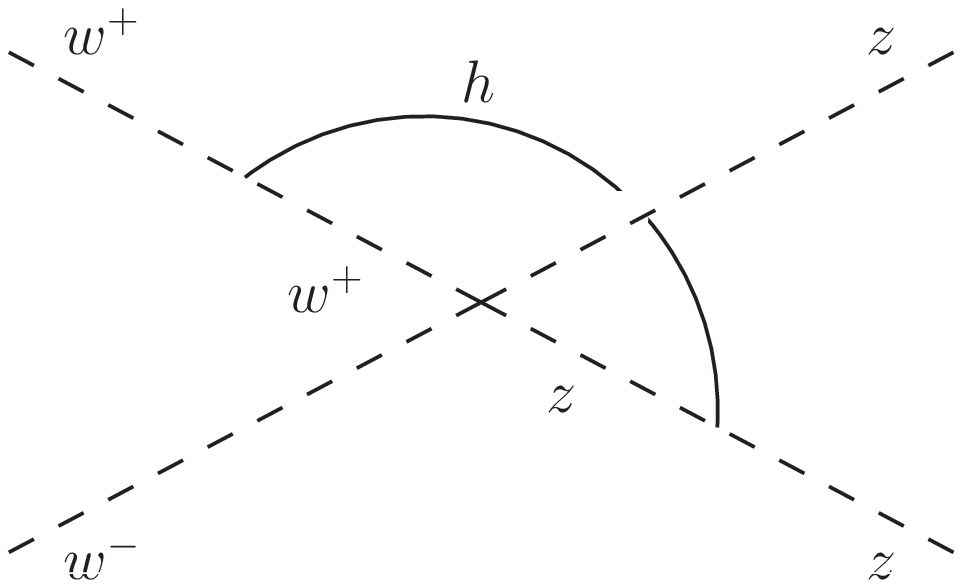}} \hspace{1cm}
\subfigure[(f)]{\includegraphics[clip,width=0.30\textwidth]{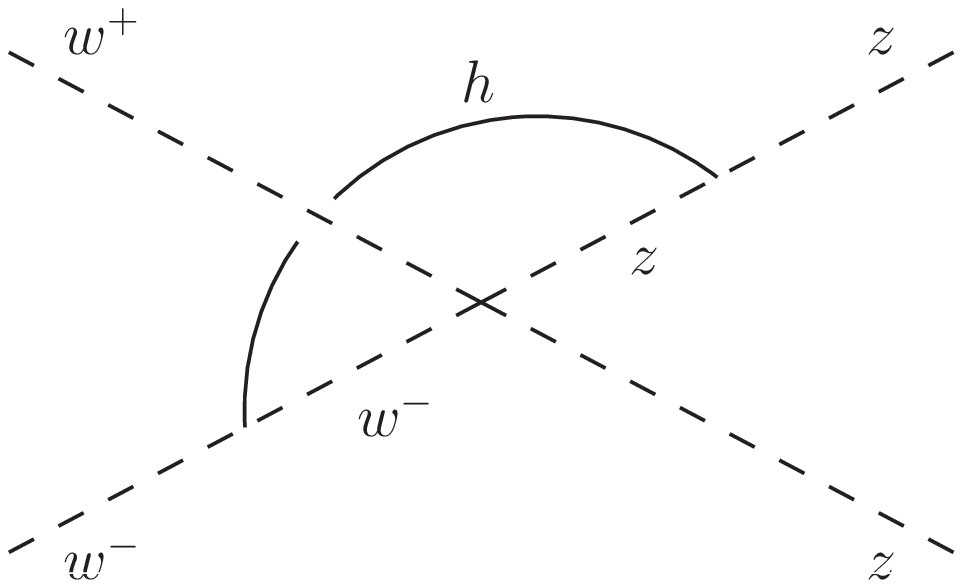}} \\
\subfigure[(g)]{\includegraphics[clip,width=0.30\textwidth]{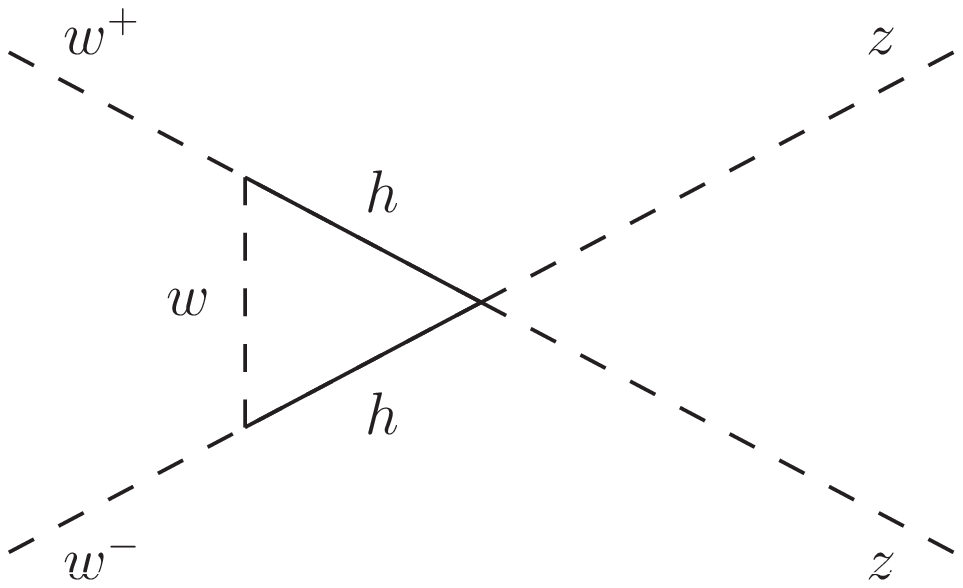}} \hspace{1cm}
\subfigure[(h)]{\includegraphics[clip,width=0.30\textwidth]{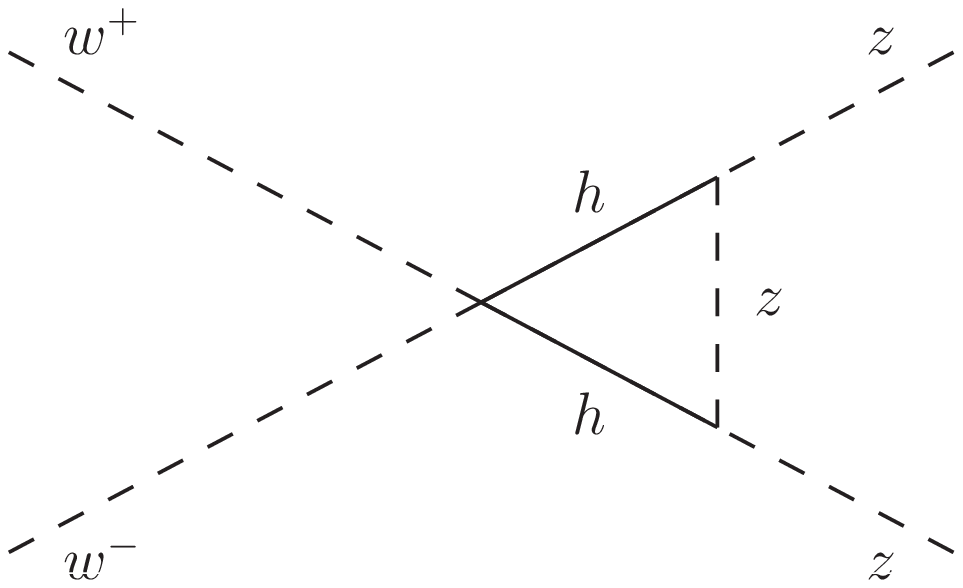}}
\caption{Triangle diagrams contributing to the irreducible part of the $w^+w^- \to zz$ amplitude, $i \mathcal{M}_{triangles}$.}
\label{fig:diagrams_triangle}
\end{figure}

The triangle diagrams are given in Fig.~\ref{fig:diagrams_triangle} and their contributions give
(for $d_{3}=d_{4}=1$) the total result
\bea
\label{eq:3.6}
\!\!\!\!\!\!\!\!\!i \mathcal{M}_{triangles} &=& 
i \lfactor \lrp{a^{2} \mhfr} \times \\ \nn
& &\!\!\!\!\!\!
\Bigg( 
\frac{2 s^{2}-t^{2}-u^{2}-18 \mhsq s}{9 \mhfr}  
- 2\lrp{2 \frac{s^{2}}{t u} - (1+b)\frac{s}{\mhsq}  - 2} \frac{A_{0}(\mhsq)}{\mhsq}
\\ \nn & &\!\!\!\!\!\!
- \frac{b\,s}{\mhsq} \fracp{s+2\mhsq}{\mhsq} B_{0}(s,\mhsq,\mhsq)
+ \frac{s}{\mhsq} \fracp{s-2\mhsq}{\mhsq} B_{0}(s,0,0)
\\ \nn & &\!\!\!\!\!\!
+ \left(\frac{1}{3} \lrp{ - \frac{t (t-u)}{\mhfr} + 3~\frac{t}{\mhsq} -  6~\frac{(2s + t)}{t} } B_{0} (t,0,0) + (t \Leftrightarrow u)\right) 
\\ \nn & &\!\!\!\!\!\!
+ 2 s \Big(b~C_{0}(0,0,s,\mhsq,0,\mhsq) - C_{0}(0,0,s,0,\mhsq,0) \Big)
\\ \nn & &\!\!\!\!\!\!
+ \left(2 \lrp{- \frac{s}{\mhsq} - \frac{(2s + t)}{t} } \mhsq ~ C_{0}(0,0,t,0,\mhsq,0)  + (t \Leftrightarrow u)\right) 
\Bigg)
\eea

\subsection{Box diagrams}

\begin{figure}[tbh]
\centering
\subfigure[(a)]{\includegraphics[clip,width=0.30\textwidth]{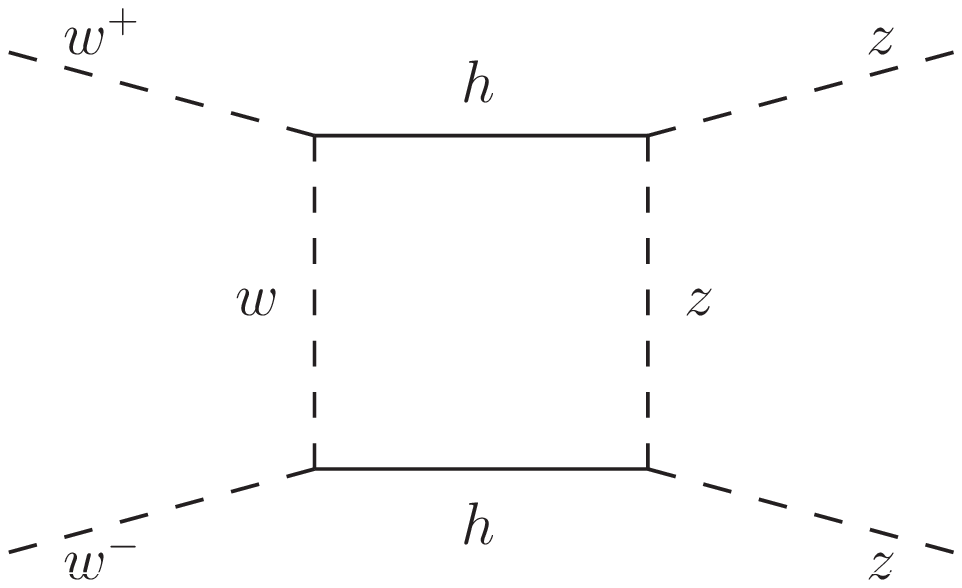}} \hspace{1cm}
\subfigure[(b)]{\includegraphics[clip,width=0.30\textwidth]{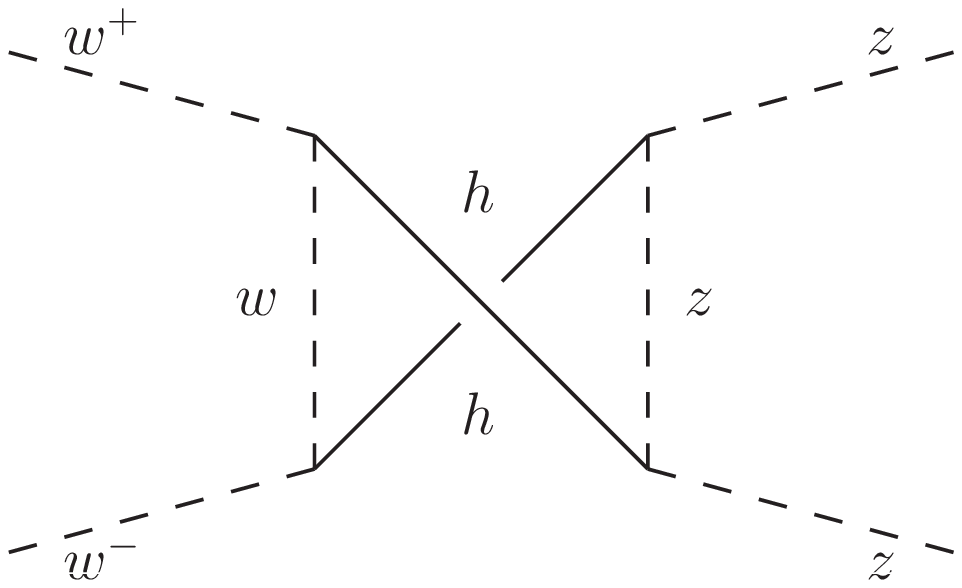}} \\
\caption{The box diagrams contributing to the irreducible part of the amplitude, $i \mathcal{M}_{boxes}$.}
\label{fig:diagrams_box}
\end{figure}

The box diagrams are depicted in Fig.~\ref{fig:diagrams_box} and their contributions 
differ only in the exchange of $t \leftrightarrow u $
\bea
\label{eq.3.8}
i \mathcal{M}_{boxes} &=& 
i \lfactor \lrp{a^{4} \mhfr} \times \\ \nn
& &
\Bigg( 
%\frac{1}{18} \fracp{- s^{2} + t^{2} + 9 \mhsq s}{\mhfr}
\frac{1}{18} \lrp{- \fracp{s}{\mhsq}^{2} + \fracp{t}{\mhsq}^{2} + 9 \fracp{s}{\mhsq}}
%\\ \nn & &
+ \lrp{-\frac{3}{2} \frac{s}{\mhsq} + 4\frac{t}{u}} \frac{A_{0}(\mhsq)}{\mhsq}
\\ \nn & &
%+ \frac{1}{4} \frac{(s + 2 \mhsq)^{2}}{\mhfr} B_{0}(s,\mhsq,\mhsq)
+ \frac{1}{4} \lrp{\frac{s}{\mhsq} + 2}^{2} B_{0}(s,\mhsq,\mhsq)
%\\ \nn & &
+ \lrp{\frac{1}{6} \frac{t(t-u)}{\mhfr} - \frac{t}{\mhsq} - 4\frac{t}{u} - 1 } B_{0}(t,0,0)
\\ \nn & &
- \lrp{\frac{s}{\mhsq} + 2} \mhsq C_{0}(0,0,s,\mhsq,0,\mhsq) + 2 \lrp{\frac{s}{\mhsq}-\frac{u}{t}} \mhsq C_{0}(0,0,t,0,\mhsq,0)
\\ \nn & &
+ \mhfr D_{0}(0,0,0,0,s,t,\mhsq,0,\mhsq,0)  \Bigg) + (t \Leftrightarrow u).
\eea
The scalar function $D_0$ is also described in the appendix.

\subsection{Five-field diagrams}

\begin{figure}[tbh]
\centering
\subfigure[(a)]{\includegraphics[clip,width=0.30\textwidth]{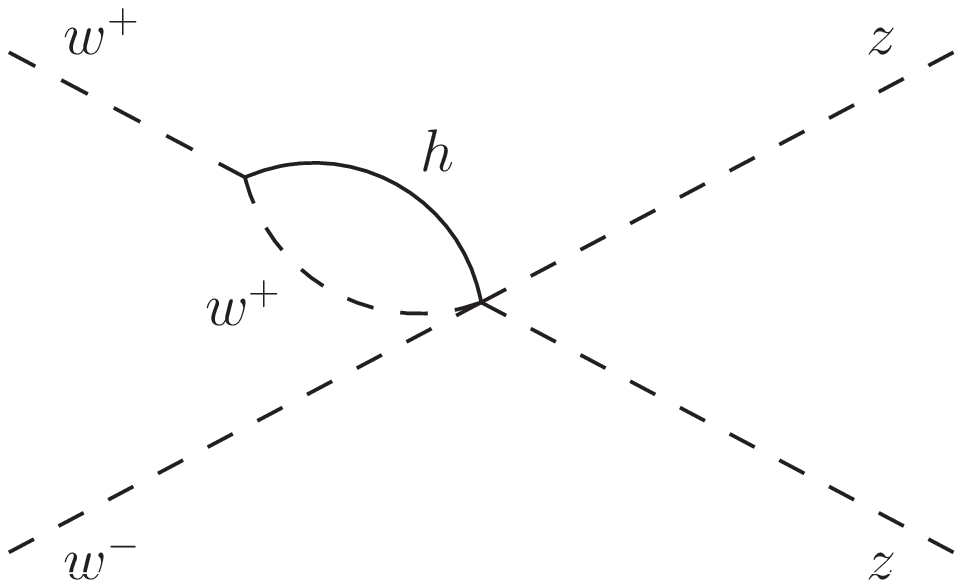}} \hspace{1cm}
\subfigure[(b)]{\includegraphics[clip,width=0.30\textwidth]{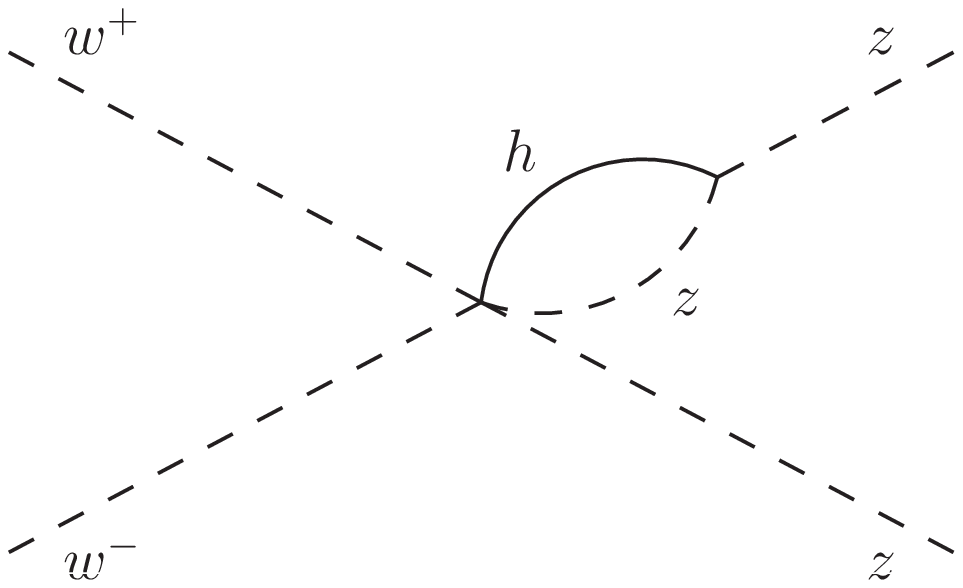}} \\
\subfigure[(c)]{\includegraphics[clip,width=0.30\textwidth]{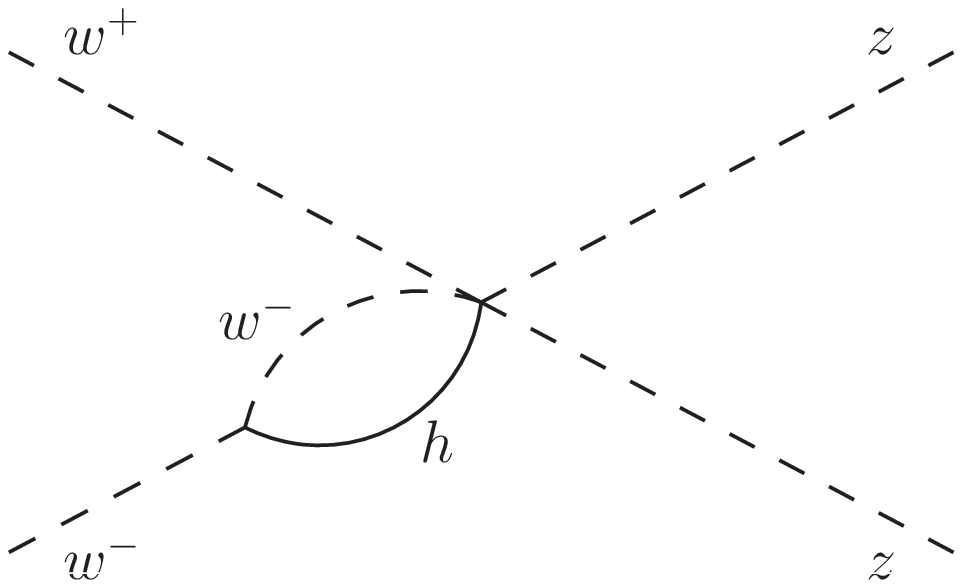}} \hspace{1cm}
\subfigure[(d)]{\includegraphics[clip,width=0.30\textwidth]{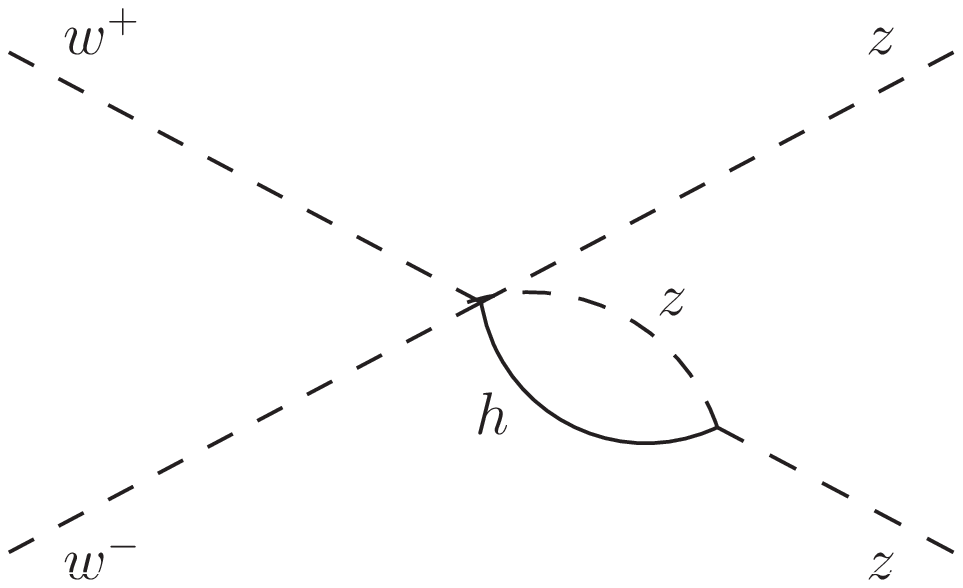}} \\
\caption{Five-field diagrams, $i \mathcal{M}_{5 F}$. Note that they do not have a counterpart in the linear realization of the SM.}
\label{fig:diagrams_five}
\end{figure}

The five-field diagrams do not have a linear calculation counterpart; they
are a new topology present in the non-linear description. They are shown 
in Fig.~\ref{fig:diagrams_five} and they are found by starting from the $wwzz$ four-point vertex and adding 
a Higgs leg to the central vertex and then connecting it to each of the four external legs. 
Their inclusion is necessary to make the calculation complete to $\mathcal{O}((M_{H}/v)^{4})$. Summed together they give
\bea
\label{eq:3.9}
i \mathcal{M}_{5 F} &=& 
i \lfactor \lrp{a^{2} \mhfr}
\fracp{s}{\mhsq} \lrp{1 + 2 \frac{A_{0}(\mhsq)}{\mhsq}}  
\eea

\subsection{Six-field diagram}

\begin{figure}[tbh]
\centering
\includegraphics[clip,width=0.30\textwidth]{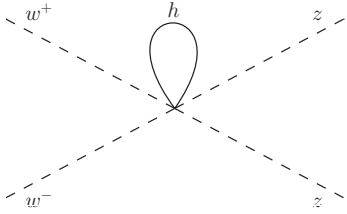}
\caption{Six-field diagram, $i\mathcal{M}_{6F}$}
\label{fig:diagrams_six}
\end{figure}

Finally, there is a single diagram here in which two Higgs legs connect to the central $wwzz$ four-point vertex and then 
connect to each other to form a single closed loop. As with the five-field case, it is again necessary to 
ensure the calculation is complete to $\mathcal{O}((M_{H}/v)^{4})$ and similarly has no linear-calculation 
counterpart. This is given in Fig.~\ref{fig:diagrams_six}. It gives
\bea
\label{eq:3.10}
i \mathcal{M}_{6F} &=& 
i \lfactor \lrp{b \mhfr} \fracp{s}{\mhsq} \fracp{-A_{0}(\mhsq)}{\mhsq}
\eea

%%%%%%%%%%%%%%%%%%%%%%%%%%%%%%%%%%%%%%%%%%%%%%%%%%%%%%%%%%%%%%%%%%%%%%%%
\section{Wave-function renormalization and tadpoles}

\subsection{Tadpoles}
\begin{figure}[tbh]
\centering
\subfigure[]  {\includegraphics[clip,width=0.30\textwidth]{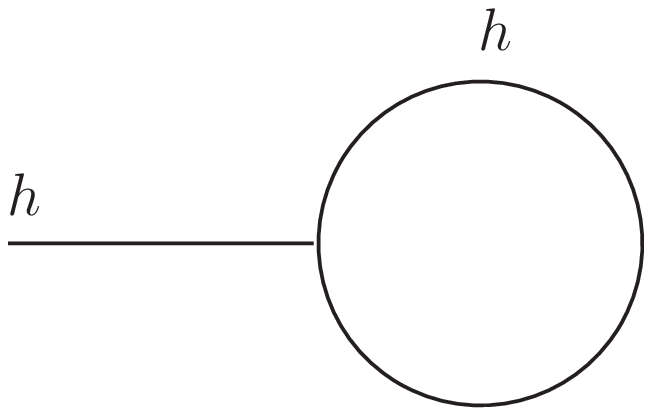}} \hspace{1cm}
\subfigure[]{\includegraphics[clip,width=0.30\textwidth]{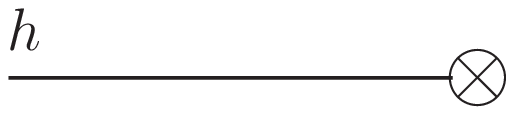}} \\
\caption{Tadpole diagram and counterterm for the Higgs field.}
\label{fig:diagrams_tadpole}
\end{figure}

The one-loop tadpole diagram and counterterm are given in Fig.~\ref{fig:diagrams_tadpole}. For $M_{w} = 0$, and 
when assuming the relationship $\lambda = \frac{\mhsq}{2 v^{2}}$ for the renormalized quantities, 
there is a single contributing diagram to the Higgs tadpole at one-loop: a Higgs loop deriving from a 
three-Higgs coupling. 
This gives a value of the tadpole (with external leg removed) of 
\bea
\label{eq:4.1}
i~T
& = & \fracp{3 d_{3} \mhsq}{2 v} \intdk \frac{1}{ \lrp{k^{2} - \mhsq}  } \\ \nn
& = & i \fracp{1}{4 \pi v^{2}}^{2} \fracp{3 \mhsq v^{3}}{2} A_{0} \lrp{\mhsq}. 
\eea
From the counterterm lagrangian Eq. (\ref{eq:1.3}) the contribution from the tadpole counterterm is
\be
\label{eq:4.2}
i~\delta T =  - i v \lrp{ \delta{\mhsq}-2 v^{2} \delta{\lambda} - 2\lambda \delta{v^{2}}}.
\ee
Therefore, to meet our renormalization condition for vanishing tadpoles at one-loop, we must have
\be
\label{eq:4.3}
\frac{\delta{\mhsq}}{\mhsq} - \frac{\delta{\lambda}}{\lambda} - \frac{\delta{v^{2}}}{v^{2}} = 
-\frac{\mhsq}{(4 \pi v)^{2} } \fracp{3}{2} \frac {A_{0} \lrp{\mhsq}}{\mhsq}
\ee

\subsection{Goldstone boson wave-function renormalization}

\begin{figure}[tbh]
\centering
{\includegraphics[clip,width=0.30\textwidth]{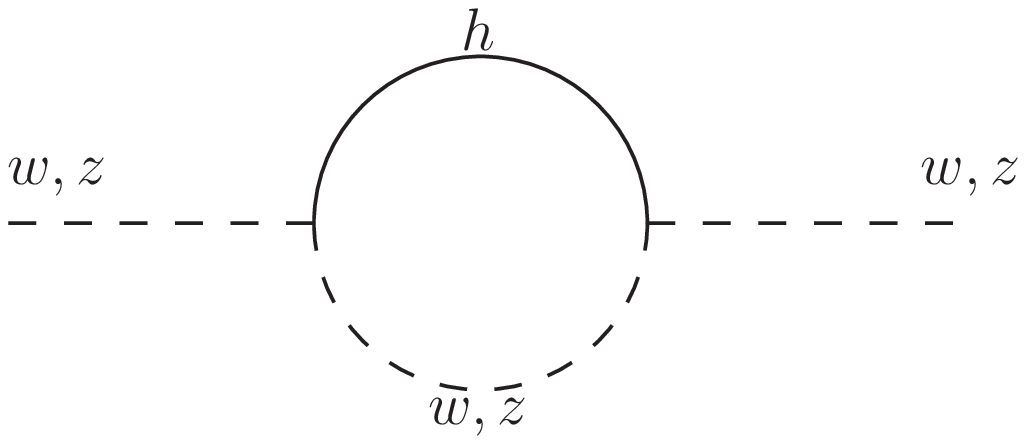}} \hspace{1cm}
{\includegraphics[clip,width=0.30\textwidth]{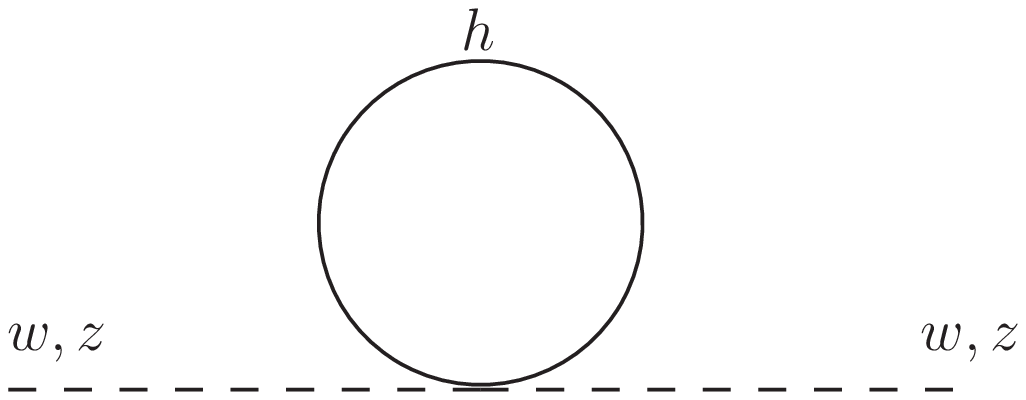}} 
\caption{Self-energy for $w$/$z$ fields (contribution to $i\mathcal{M}_{{\text WFR}} $)}
\label{fig:diagrams_two-point}
\end{figure}
When all Higgs tadpoles are appropriately canceled, there are only mixed Higgs/Goldstone boson loops, a Higgs loop, 
and $w/z$ loops (which are zero when the $w/z$ are massless). Any divergences which appear due to the wave-function 
renormalization of the external fields must be canceled by something in the remainder of this amplitude. 
We shall see later that this is easily achieved with the renormalization of $v^2$, which is 
also a global factor multiplying the tree-level contribution. In fact from
the mere requirement of finiteness of the amplitude after including the one loop diagrams, we can derive only
a condition on the combination $2\delta Z_w - \delta v^2$. Therefore the renormalization condition on the 
wave function has to be imposed separately and this consists in requesting the unit residue condition on the external legs.

The two-point function for the Goldstone bosons in Fig.~\ref{fig:diagrams_two-point} gives the following
\bea
\label{eq:4.4}
- i \Pi_{w}(q^{2}) \equiv - i \Pi_{w}^{(1+2)}(q^{2}) &=&  i \fracp{1}{4 \pi v^{2}}^{2} (v^{2})
\Big( \lrp{a^{2} (3 q^{2} - \mhsq)- b q^{2}}  A_{0}(\mhsq) \\ \nn
& &  \hspace{2.75cm} + a^{2} (q^{2}-\mhsq)^{2} B_{0}(q^{2},0,\mhsq) \Big),
\eea
which verifies $\Pi_{w}(0) = 0$ for all $a$ and $b$, and therefore the Goldstone bosons stay massless, 
as they should\footnote{To see this it is important to note that $B_{0}(0,0,\mhsq) = A_{0}(\mhsq)/\mhsq$.}. 

The wavefunction renormalization factor is then
\bea
\label{eq:4.5}
Z_{w}
&=& 1 + \left. \frac{d \Pi_{w}}{d q^{2}} \right|_{q^{2}=0} 
= 1 - \fracp{1}{4 \pi v}^{2} \lrb{ (a^{2}-b) A_{0}(\mhsq) + a^{2} \frac{\mhsq}{2}} 
\eea
In the SM case, this is finite and matches the value given by ref.~\cite{gupta}
\bea
\label{eq:4.6}
Z_{w}^{\rm SM} 
&=& \lrb{1 - \fracp{1}{4 \pi v}^{2} \lrp{ \frac{\mhsq}{2}}} 
= \lrb{1 - \frac{\lambda^{2}}{16 \pi^{2}} } 
\eea
but in general it is divergent. This divergence is canceled against contributions from $\delta{v^{2}}$ 
when the corresponding contribution to the one-loop amplitude is placed in the complete calculation. The one-loop
contribution to the amplitude $w^+ w^-\to zz$ from wave-function renormalization is
\be
\label{eq:4.7}
i \mathcal{M}_{\rm WFR} 
= i \fracp{1}{4 \pi v^{2}}^{2} \lrp{\mhsq s} \fracp{(a^{2}-1)s + \mhsq}{s-\mhsq} \lrp{ a^{2} + 2(a^{2}-b) \frac{A_{0}(\mhsq)}{\mhsq} }
\ee

\subsection{Higgs boson wave-function renormalization}
 
The contributions to the Higgs two-point function can be derived from Sec.~\ref{sec:two_point}, 
while the counterterm contribution is simply
\be
\label{4.8}
- i \Pi_{H}^{(ctr)}(q^{2}) = - i \delta{\mhsq} 
\ee
This gives
\bea
\label{eq:4.9}
- i \Pi_{H}(q^{2}) &=& i \fracp{1}{4 \pi v}^{2}
\fracp{3 \mhfr}{2} \Bigg( \frac{d_{4}}{\mhsq} A_{0}(\mhsq) + 3 d_{3}^{2} B_{0}(q^{2},\mhsq,\mhsq) \\ \nn 
& & \hspace{3.75cm} + a^{2} \frac{s^{2}}{\mhfr} B_{0}(q^{2},0,0) \Bigg)
 - i \delta{\mhsq} 
\eea
The on-shell condition for the Higgs mass requires 
\be
\label{eq:4.10}
{\rm Re} \, \Pi_{H}(\mhsq) = 0
\ee
Independent of this condition and the counterterm, we have the wavefunction renormalization factor 
of (now setting $d_{3}=d_{4}=1$)
\bea
\label{eq:4.11}
Z_{H}
&=& 1 + \left. \frac{d \Pi_{H}}{d q^{2}} \right|_{q^{2}=\mhsq} \\ \nn
&=& 1 - \fracp{1}{4 \pi v}^{2} \fracp{3 \mhfr}{2}
\lrb{ 3 B_{0}'(\mhsq,\mhsq,\mhsq) + a^{2} \lrp{  B_{0}'(\mhsq,0,0) + \frac{2}{\mhsq} B_{0}(\mhsq,0,0)} } \\ \nn
&=& 1 + \fracp{M_{H}}{4 \pi v}^{2}
\lrb{ \fracp{9}{2}\lrp{1-\frac{2 \sqrt{3} \pi}{9}} + a^{2} \lrp{\frac{3}{2} - B_{0}(\mhsq,0,0)}}
\eea
This is divergent in the SM case and only becomes finite for $a=0$. When the one-loop correction 
to $i \Gamma(h \to w^{+} w^{-})$ is performed and all external wavefunction renormalizations are included 
(i.e. both $Z_{H}$ and $Z_{w}$), all divergences cancel for arbitrary $a$ and $b$ when using the appropriate 
values for the counterterms given in Section \ref{sec:divergences}. This is a good check on this 
value of $Z_{H}$. It should also be noted that the SM value for $Z_{H}$ does not match that given in ref.~\cite{gupta}; 
this is a result of the nonlinear nature of the calculation.

The complete, renormalized decay width for the Higgs boson into Goldstone bosons is
\bea
\label{eq:4.12}
\Gamma(h \to w w) &=& \fracp{3 \lambda M_{H} a}{16 \pi} \Bigg( a + \fracp{\lambda}{\pi^{2}} 
\Bigg\{  \fracp{1}{16} \Big( a \big( 17+10b-3a(7a-8)  \\ \nn
& & \hspace{6.5cm}
-2~\overline{\delta v^2} \big) -12b \Big)  \\ \nn
& & \hspace{2cm} - \fracp{\pi}{8} \sqrt{3} \lrp{1+3a-b} + \fracp{\pi^{2}}{48} (a^{2})\lrp{4+a}
\Bigg\} \Bigg),
\eea
for arbitrary $a$ and $b$,
where $\overline{\delta v^2}$ is a finite renormalization, not fixed by our conditions. 
For $a=b=1$ and $~ \overline{\delta v^2}=-\frac{1}{2}$ (the value used in refs.~\cite{DW} and~\cite{gupta}), this reproduces the known SM result
\bea
\label{eq:4.13}
\Gamma(h \to w w) &=& \fracp{3 \lambda M_{H}}{16 \pi} \Bigg( 1 + \fracp{\lambda}{\pi^{2}} 
\Bigg\{  \frac{19}{16} - \frac{3 \sqrt{3} \pi}{8} + \frac{5\pi^{2}}{48} 
\Bigg\} \Bigg)
\eea

%%%%%%%%%%%%%%%%%%%%%%%%%%%%%%%%%%%%%%%%%%%%%%%%%%%%%%%%%%%%%%%%%%%%%%%%%%%%%%%%%%%%%%%
\section{Divergences and determination of the counterterms}\label{sec:divergences}
Here we give the pieces of each individual diagram proportional to 
$\Delta_{\epsilon} = \fracp{2}{\epsilon} - \gamma_{E} + \log4\pi + \log\frac{\mu^{2}}{\mhsq}$.
We give the results in the case $d_3=d_4=1$ but it is quite straightforward to restore
these factors for each individual diagram if so desired. These factors appear only in
the radiative corrections to two- and three-point functions.
\bea
\mathcal{M}_{2-pt}^{(a)} & \sim & \fracp{1}{4 \pi v^{2}}^{2} \Delta_{\epsilon} \fracb{9 a^{2} \mhfr s^{2}}{2(s-\mhsq)^{2}}
\\ \nn
\mathcal{M}_{2-pt}^{(b)} & \sim & \fracp{1}{4 \pi v^{2}}^{2} \Delta_{\epsilon} \fracb{3 a^{4} s^{4}}{2(s-\mhsq)^{2}}
\\ \nn
\mathcal{M}_{2-pt}^{(c)} & \sim & \fracp{1}{4 \pi v^{2}}^{2} \Delta_{\epsilon} \fracb{3 a^{2} \mhfr s^{2}}{2(s-\mhsq)^{2}}
\eea

\bea
\mathcal{M}_{3-pt}^{(a)} & \sim & \fracp{1}{4 \pi v^{2}}^{2} \Delta_{\epsilon} \fracb{3 a b \mhsq s^{2}}{(s-\mhsq)}
\\ \nn
\mathcal{M}_{3-pt}^{(b)} & \sim & \fracp{1}{4 \pi v^{2}}^{2} \Delta_{\epsilon} \fracb{-2 a^{2} s^{3}}{(s-\mhsq)}
\\ \nn
\mathcal{M}_{3-pt}^{(c)} & \sim & \fracp{1}{4 \pi v^{2}}^{2} \Delta_{\epsilon} \fracb{-2 a^{2} b \mhsq s^{2}}{(s-\mhsq)}
\\ \nn
\mathcal{M}_{3-pt}^{(d)} & \sim & \fracp{1}{4 \pi v^{2}}^{2} \Delta_{\epsilon} \fracb{-3 a^{3} \mhsq s^{2}}{(s-\mhsq)}
\\ \nn
\mathcal{M}_{3-pt}^{(e)} & \sim & \fracp{1}{4 \pi v^{2}}^{2} \Delta_{\epsilon} \fracb{-a^{4} s^{2} \lrp{s-2\mhsq}}{(s-\mhsq)}
\eea

\bea
\mathcal{M}_{bubble}^{(a)} & \sim & \fracp{1}{4 \pi v^{2}}^{2} \Delta_{\epsilon} \fracb{b^{2} s^{2}}{2}
\\ \nn
\mathcal{M}_{bubble}^{(b)} & \sim & \fracp{1}{4 \pi v^{2}}^{2} \Delta_{\epsilon} \fracb{s^{2}}{2}
\\ \nn
\mathcal{M}_{bubble}^{(c)} & \sim & \fracp{1}{4 \pi v^{2}}^{2} \Delta_{\epsilon} \fracb{t(t-u)}{6}
\\ \nn
\mathcal{M}_{bubble}^{(d)} & \sim & \fracp{1}{4 \pi v^{2}}^{2} \Delta_{\epsilon} \fracb{u(u-t)}{6}
\eea

\bea
\mathcal{M}_{triangle}^{(a)+(b)} & \sim & \fracp{1}{4 \pi v^{2}}^{2} \Delta_{\epsilon} \fracb{a^{2} s (3s - 2\mhsq)}{3}
\\ \nn
\mathcal{M}_{triangle}^{(c)+(d)} & \sim & \fracp{1}{4 \pi v^{2}}^{2} \Delta_{\epsilon} \fracb{-a^{2} t ((t-u)-\mhsq)}{3}
\\ \nn
\mathcal{M}_{triangle}^{(e)+(f)} & \sim & \fracp{1}{4 \pi v^{2}}^{2} \Delta_{\epsilon} \fracb{-a^{2} u ((u-t)-\mhsq)}{3}
\\ \nn
\mathcal{M}_{triangle}^{(g)+(h)} & \sim & \fracp{1}{4 \pi v^{2}}^{2} \Delta_{\epsilon} \lrb{- a^{2} b s^{2}}
\eea

\bea
\mathcal{M}_{box}^{(a)+(b)} & \sim & \fracp{1}{4 \pi v^{2}}^{2} \Delta_{\epsilon} \fracb{a^{4}\lrp{s^{2}+t^{2}+u^{2}}}{3}
\eea

\bea
\mathcal{M}_{5F}^{(a)+(b)+(c)+(d)} & \sim & \fracp{1}{4 \pi v^{2}}^{2} \Delta_{\epsilon} \lrb{2 a^{2} \mhsq s}
\eea

\bea
\mathcal{M}_{6F} & \sim & \fracp{1}{4 \pi v^{2}}^{2} \Delta_{\epsilon} \lrb{-b \mhsq s}
\eea

\bea
\mathcal{M}_{WFR} & \sim & \fracp{1}{4 \pi v^{2}}^{2} \Delta_{\epsilon} \lrp{2 \mhsq s }(a^{2}-b)\fracp{(a^{2}-1)s+\mhsq}{(s-\mhsq)}
\eea
Note that we have included the $w,z$ Goldstone boson wave-function renormalization as a contribution to the one-loop amplitude
to be canceled by the counterterms in $\delta {\mathcal L}$.

If we ignore the tadpole counterterms, we can collect together all the individual counterterms to give the following
\bea
\label{eq:amp_counterterm}
\mathcal{M}^{{\text ctr}} &=& 
\fracp{1}{v^{2}}\fracp{s}{\lrp{s-\mhsq}^{2}} \Bigg(
\frac{\delta{v^{2}}}{v^{2}} \lrb{ (a^{2}-1)s^{2} + (2-a^{2})(s \mhsq) - \mhfr } 
\\ \nn & & \hspace{3.25cm}
 - \delta{a} \lrb{(2a)(s)(s-\mhsq)}  - \frac{\delta{\mhsq}}{\mhsq} \lrb{(a^{2})(s \mhsq)}
\Bigg)
\\ \nn & & \hspace{0.25cm}
+ \fracp{1}{v^{4}} \Bigg(
4 \delta{a_{4}} (t^{2}+u^{2}) + 8 \delta{a_{5}} (s^{2}) 
\Bigg)
\eea
The values of the counterterms needed to cancel the one-loop divergences---and satisfy our renormalization conditions---
can be solved for arbitrary $a$ and $b$ to give
\bea
\label{eq:counterterms_solution}
\frac{\delta{v^{2}}}{v^{2}} &=& \frac{\mhsq}{\lrp{4\pi v}^{2}} \lrb{\Delta_{\epsilon} (-a^{2}+b) + \overline{\delta v^2}}\\ \nn
\frac{\delta{\mhsq}}{\mhsq}
&=& \frac{\mhsq}{\lrp{4\pi v}^{2}}  \fracp{3}{2}\lrb{ \Delta_{\epsilon}\lrp{4+a^{2}} + 7+2a^{2} -\sqrt{3} \pi}\\ \nn 
\frac{\delta{\lambda}}{\lambda}  &=& \frac{\mhsq}{\lrp{4\pi v}^{2}}  \fracp{1}{2}\lrb{ \Delta_{\epsilon}\lrp{9+5a^{2}-2b} + 18+6a^{2} -3\sqrt{3} \pi - 2 \overline{\delta v^2}}\\ \nn
\delta{a}      &=& \frac{\mhsq}{\lrp{4\pi v}^{2}} \fracp{1}{2} \lrb{ \Delta_{\epsilon} (a-1) (a (5 a+2)-3b) } \\ \nn 
\delta{a_{4}}   &=&  \frac{1}{\lrp{4\pi}^{2}} \fracp{-1}{12}\lrb{ \Delta_{\epsilon}\lrp{a^{2}-1}^{2}  } \\ \nn
\delta{a_{5}}   &=&  \frac{1}{\lrp{4\pi}^{2}} \fracp{-1}{48} \lrb{ \Delta_{\epsilon} \lrp{2+5a^{4}-4a^{2}-6a^{2}b+3b^{2}} }\\ \nn
\eea
where $\overline{\delta v^2}$ is a finite piece, not determined by the renormalization conditions a priori.
Note that the counterterm for $b$ cannot determined from this process. As previously indicated it is 
quite easy to restore the 
dependence on $d_3$ 
and $d_4$ in the divergent part of all diagrams but we will not present the results here. 

\subsection{Cross-checks}

In the SM case ($a=b=1$), renormalization conditions read as
\bea
\frac{\delta{v^{2}}}{v^{2}} &=& \frac{\mhsq}{\lrp{4\pi v}^{2}} \lrb{\overline{\delta v^2}}\\ \nn
\frac{\delta{\mhsq}}{\mhsq}
&=& \frac{\mhsq}{\lrp{4\pi v}^{2}}  \fracp{3}{2}\lrb{ \Delta_{\epsilon}(5) + 9 -\sqrt{3} \pi}\\ \nn
\frac{\delta{\lambda}}{ \lambda} &=& \frac{\mhsq}{\lrp{4\pi v}^{2}} \lrb{ \Delta_{\epsilon}(6) + 12 -\frac{3}{2}\sqrt{3} \pi - \overline{\delta v^2} }\\ \nn
\delta{a}      &=& 0 \\ \nn 
\delta{a_{4}}   &=& 0 \\ \nn
\delta{a_{5}}   &=& 0 \nn
\eea
The last three terms should be absent in the SM, so this is a good check.  In the EChL case ($a=b=0$) we have
\bea
\frac{\delta{v^{2}}}{v^{2}} &=& \frac{\mhsq}{\lrp{4\pi v}^{2}} \lrb{\overline{\delta v^2}}\\ \nn
\frac{\delta{\mhsq}}{\mhsq}
&=& \frac{\mhsq}{\lrp{4\pi v}^{2}} \fracp{3}{2} \lrb{ \Delta_{\epsilon}(4) + 7 -\sqrt{3} \pi}\\ \nn
\frac{\delta{\lambda}}{\lambda} &=& \frac{\mhsq}{\lrp{4\pi v}^{2}} \lrb{ \Delta_{\epsilon}\fracp{9}{2} + 9 -\frac{3}{2}\sqrt{3} \pi - \overline{\delta v^2} }\\ \nn
\delta{a}      &=& 0 \\ \nn 
\delta{a_{4}}   &=& \frac{1}{\lrp{4\pi}^{2}}  \Delta_{\epsilon}  \fracp{-1}{12} \\ \nn
\delta{a_{5}}   &=& \frac{1}{\lrp{4\pi}^{2}}  \Delta_{\epsilon}  \fracp{-1}{24} \nn
\eea
in agreement with already known results \cite{ECHL}.

It is interesting to note here that while $\delta{\mhsq} \ne 0$, its contribution to the counterterm 
amplitude is actually proportional to $a^{2}$ and therefore vanishes when $a \to 0$ (see Eq.~\ref{eq:amp_counterterm}). 
Also, the $\delta{\lambda}$ term is only necessary here to remove the tadpole divergence 
(which is absent from the full amplitude for $a=b=0$), so once again plays no part. 
Finally, the $\delta{v^{2}}$ term is finite. Therefore only $\delta{a_{4}}$ and $\delta{a_{5}}$ are needed to 
remove the one-loop divergences from the Goldstone boson scattering amplitudes, which is what one would expect
in the EChL approach.

%%%%%%%%%%%%%%%%%%%%%%%%%%%%%%%%%%%%%%%%%%%%%%%%%%%%%%%%%%%%%%%%%%%%%%%%%%%%%%%%%%%%%%%%%%%%%%%%%%%%%
\section{Final result and conclusions}
Finally, the complete one-loop amplitude $i\mathcal{M}^{loop} (w^{+}w^{-} \to z z)$ (for arbitrary $a$ and $b$ 
and rendered finite by using the counterterms in Eq.~\ref{eq:counterterms_solution}) is given by the following
\bea
\label{eq:final_amplitude}
i \mathcal{M}^{loop}  &=&
i \lfactor \fracp{\mhsq}{2}^{2} \Bigg(
   \frac{6 a^{2}(-6-2a^{2}+\sqrt{3}\pi) \mhfr}{(s-\mhsq)^{2}}
\\ \nn & &\!\!\!
- \frac{4 a^{2} \lrp{ 18+2a(a-3)+5b-3\sqrt{3}\pi - \overline{\delta v^2} }\mhsq}{(s-\mhsq)}
\\ \nn & &\!\!\!
- \frac{2}{9} (a^{2}-1) \lrb{(a^{2}-1)\, \frac{t^{2}+4tu+u^{2}}{\mhfr} - 72 a^{2}\, \frac{t^{2}+tu+u^{2}}{t u}}
\\ \nn & &\!\!\!
+ 4 \fracp{s}{\mhsq} \lrb{2a^{4}-3a^{2}b+b+(a^{2}-1)\overline{\delta v^2} } 
+ a^{2} \lrb{6 \sqrt{3} \pi-4(9+3a(a-2)+5b-\overline{\delta v^2}\,)}
\\ \nn & &\!\!\! 
+ 2 \fracb{(a^{2}-b)s^{2} + ((a^{2}-b)-3a)\mhsq s -2 a^{2} \mhfr}{\mhsq (s-\mhsq)}^{2} \overline{B}_{0}(s,\mhsq,\mhsq)
\\ \nn & &\!\!\!
+ 2 \frac{ s\lrb{ (a^{2}-1)s + \mhsq } }{\mhfr}
\fracb{(a^{2}-1)s^{2} + (6a^{2}+1) \mhsq s - 4 a^{2} \mhfr}{(s-\mhsq)^{2}} \overline{B}_{0}(s,0,0)
\\ \nn & &\!\!\!
+ \Bigg( \Bigg(\frac{4(a^{2}-1)^{2}}{3} \frac{t^{2}}{\mhfr} + \frac{2(a^{2}-1)}{3} \frac{((a^{2}-1)s-6 a^{2}\mhsq)}{\mhsq}\frac{t}{\mhsq} 
\\ \nn & &\!\!\! \hspace{1cm}
+ 4 a^{2} \lrp{1-(a^{2}-1)\lrp{1+4\frac{u}{t} }} \Bigg)
\overline{B}_{0}(t,0,0)  +  (t \Leftrightarrow u) \Bigg)
\\ \nn & & \\ \nn & &
\!\!\!- 8 a^{2} \frac{(a^{2}-b)s^{2} + ((a^{2}-b)-3a)\mhsq s -2 a^{2} \mhfr}{(s-\mhsq)}\, C_{0}(0,0,s,\mhsq,0,\mhsq)
\\ \nn & &
\!\!\!+ 8 a^{2} \frac{s}{(s-\mhsq)} \lrb{(a^{2}-1)s+\mhsq} C_{0}(0,0,s,0,\mhsq,0)
\\ \nn & &
+ \Bigg( 8 a^{2} \lrb{ (a^{2}-1)s + \mhsq \lrp{1 - 2(a^{2}-1)\,\frac{u}{t} } }  C_{0}(0,0,t,0,\mhsq,0)
+ (t \Leftrightarrow u) \Bigg)
\\ \nn & &  \\ \nn & &
\!\!\!+ \Bigg( 4 a^{2} \mhfr D_{0}(0,0,0,0,s,t,\mhsq,0,\mhsq,0) + (t \Leftrightarrow u) \Bigg)
\Bigg)
\eea
Here the functions $\bar{A_{0}}$ and $\bar{B_{0}}$ 
are the corresponding scalar integral functions with the divergences removed (see appendix). 
The amplitude as written above has been grouped by scalar loop integrals. In the SM limit ($a=b=1$), 
this simplifies quite a bit
\bea
i \mathcal{M}^{loop}_{SM}  &=&
i \lfactor \fracp{\mhsq}{2}^{2} \Bigg(
\\ \nn & &
+ \frac{\mhfr}{(s-\mhsq)^{2}} \lrb{ -48 + 6 \sqrt{3} \pi + 18 \overline{B}_{0}(s,\mhsq,\mhsq) + 6 \overline{B}_{0}(s,0,0) }
\\ \nn & &
+ \frac{\mhsq}{(s-\mhsq)} \Big( -76 +12\sqrt{3}\pi + 4\overline{\delta v^2} + 12 \overline{B}_{0}(s,\mhsq,\mhsq) + 20 \overline{B}_{0}(s,0,0)
\\ \nn & & \hspace{2.5cm}
+ (8 \mhsq) \Big( 3 C_{0}(0,0,s,\mhsq,0,\mhsq) + C_{0}(0,0,s,0,\mhsq,0) \Big) \Big)
\\ \nn & &
+ \Big( 2 \overline{B}_{0}(s,\mhsq,\mhsq) + 14 \overline{B}_{0}(s,0,0)
+ 4 \overline{B}_{0}(t,0,0) + 4 \overline{B}_{0}(u,0,0) \Big)
\\ \nn & &
+ (8\mhsq) \Big( C_{0}(0,0,s,\mhsq,0,\mhsq) +  C_{0}(0,0,s,0,\mhsq,0)
\\ \nn & & \hspace{1.5cm}
 + C_{0}(0,0,t,0,\mhsq,0) + C_{0}(0,0,u,0,\mhsq,0) \Big)
\\ \nn & &
+ (4 \mhfr) \Big( D_{0}(0,0,0,0,s,t,\mhsq,0,\mhsq,0) +  D_{0}(0,0,0,0,s,u,\mhsq,0,\mhsq,0) \Big)
\\ \nn & &
-2 \lrp{22 - 3\sqrt{3} \pi -2 \overline{\delta v^2} }
\Bigg)
\eea
Eqs. \ref{eq:counterterms_solution} and \ref{eq:final_amplitude} contain our main results. We have seen how the
departures from the SM can be taken consistently into account in an effective-Lagrangian philosophy also 
at the one-loop level and the suitable counterterms included to render the amplitude finite. We note that
if $a$ and $b$ are set to their SM values, the coefficients accompanying the $O(p^4)$ operators are finite
and do not run, while this is not the case as soon as one departs from the SM. After cancellation
of the divergent part of the loop (say in the $\overline{MS}$ scheme), a finite logarithmic part remains. For
instance in the case of the effective coefficients $a_4$ and $a_5$, and appealing to naturality arguments, their
characteristic size would be
\bea
\label{eq:estimate}
\delta{a_{4}}   &=&  \frac{1}{\lrp{4\pi}^{2}} \fracp{-1}{12}\lrp{a^{2}-1}^{2} \log\frac{f^2}{v^2} \\ \nn
\delta{a_{5}}   &=&  \frac{1}{\lrp{4\pi}^{2}} \fracp{-1}{48}\lrp{2+5a^{4}-4a^{2}-6a^{2}b+3b^{2}} \log\frac{f^2}{v^2}  \nn
\eea
$f$ being the compositeness scale.

In the present study we have restricted ourselves to the case where the triple and quartic Higgs coupling take the
same values as in the SM, but relaxing this hypothesis is straightforward. The dependence of the divergent parts on 
$d_3$ and $d_4$ can be easily determined as they simply contribute as overall factors to vertex and self-energy 
corrections.
None of those diagrams behave as $\sim s^2$ (or as $t^2$ or $u^2$) for large values of $s$ and 
they therefore do not contribute to
$\delta a_4$ and $\delta a_5$ that are totally independent of $d_3$ and $d_4$.

It would be interesting to extend the present study to other low energy constants of the effective 
theory parametrizing 
the EWSBS. In particular $a_1$ and $a_2$ correspond to operators that contribute to the triple gauge boson vertex 
that has been recently measured for the first time at the LHC \cite{TGC}. The renormalization of $d_3$ and $d_4$
would eventually be of interest too, but their relevance for comparison with experiment is still well ahead.

We have also presented a full one-loop calculation using the Equivalence Theorem approximation (and taking the 
masses of the Goldstone bosons to vanish, i.e. in the 't Hooft-Landau gauge)  of the $W_LW_L \to Z_LZ_L$ in the general case
with generic couplings of the Higgs to the electroweak gauge bosons. This calculation should be quite useful in
precise comparisons of measurements of the four gauge boson coupling (not yet measured at the LHC) to theoretical
predictions.  Its knowledge is also very relevant in connection with unitarity analysis such as the one done in \cite{EY}
and the prediction of new resonances originating from the EWSBS. As emphasized in the introduction, the search for such 
resonances has to go hand-in-hand with accurate measurements of the four gauge boson couplings. 
Almost any deviation of these coefficients 
from their SM values would lead to unitarity violations at high energies and thus require additional 
resonances to restore it.
In a forthcoming publication we will study in detail the issue of unitarity and extend the results of \cite{EY} to the case
where the tree-level $O(p^2)$ parameters $a$ and $b$ depart from their SM values. Both the determination of the counterterms
and the full calculation of the real part of the scattering amplitude derived in this preparatory paper are 
necessary ingredients for such
an analysis.  
   
In conclusion, we have successfully provided a one-loop theory of Goldstone boson scattering in the context of
an extended EWSBS where the Higgs is allowed to have arbitrary couplings. The coefficients $a$ and $b$ describing the
coupling of the Higgs to the $W$ and $Z$ gauge bosons are currently of great interest to SM fits but their 
treatment so far has only been of tree-level studies. If $a$ and $b$ are not exactly equal to one some $O(p^4)$ operators
with  running coefficients are required for a consistent treatment at one loop. Their running has been determined in 
this work. The results smoothly connect to the SM and are, we believe, completely general.

%%%%%%%%%%%%%%%%%%%%%%%%%%%%%%%%%%%%%%%%%%%%%%%%%%%%%%%%%%%%%%%%%%%%%%%%%%%%%%%%%%%%%%%%%%%%
\section*{Acknowledgements}
We gratefully acknowledge the financial support 
from projects FPA2010-20807, 2009SGR502 and CPAN (Consolider CSD2007-00042).
We thank A. Pomarol for discussions.

\section*{Appendix}

Here we define the independent scalar integrals entering our expressions
\bea
A_0(m^2_0) &=& {\cal N}\int d^d k
\dfrac{1}{k^2-m^2_0}=m^2_0\left(\Delta_{\epsilon}+1\right)\\ \nn
B_0(p_1^2,m^2_0,m^2_1) &=& 
{\cal N}\int d^d k 
\dfrac{1}{k^2-m^2_0}\dfrac{1}{(k+p_1)^2-m^2_1}\\ \nn
C_0(p_1^2,p_2^2,p_{12}^2,m^2_0,.,m^2_3) &=& 
{\cal N}\int d^d k 
\dfrac{1}{k^2-m^2_0}\dfrac{1}{(k+p_1)^2-m^2_1}
\dfrac{1}{(k+p_{12})^2-m^2_2}\\ \nn
\eea
\be
D_0(p_1^2,p_2^2,p_3^2,p_{13}^2,p_{12}^2,
p_{23}^2,m^2_0,.,m^2_3)
= 
\ee
\be
{\cal N}
\int d^d k 
\dfrac{1}{k^2-m^2_0}\dfrac{1}{(k+p_1)^2-m^2_1}
\dfrac{1}{(k+p_{12})^2-m^2_2}
\dfrac{1}{(k+p_{13})^2-m^2_3}
\ee
where ${\cal N}=(2\pi\mu)^{4-d}/(i\pi^2)$ and $p_{ij}=\sum_{h=i}^j p_h$. 
We note that of the scalar loop integrals  ($A_{0}$, $B_{0}$, $C_{0}$, and $D_{0}$) in our solution only $A_{0}$ and $B_{0}$ 
contain divergences. We will therefore define the functions $\bar{A_{0}}$ and $\bar{B_{0}}$ 
as the corresponding scalar integral functions with the divergences removed
\bea
A_{0}(a) &=& a\Delta_{\epsilon} + \bar{A_{0}}(a) \\ \nn
B_{0}(a,b,c) &=& \Delta_{\epsilon} + \bar{B_{0}}(a,b,c) \nn
\eea
Note that this differs slightly from the $\Delta_{\epsilon} = (\frac{2}{\epsilon} - \gamma_{E} + \log4\pi)$ 
used in the literature on the scalar loop integrals. However, this has the benefit that 
all factors of $\log\frac{\mu^{2}}{\mhsq}$ are currently in the counterterms and that
$\bar{A_{0}}(\mhsq) = \mhsq$.
For situations in which it is better to have $\log\frac{\mu^{2}}{\mhsq}$ explicitly in the amplitude 
(for instance in the limit $\mhsq \to \infty$), this can be achieved by replacing each 
counterterm in Eq.~\ref{eq:amp_counterterm} with $\mathcal{C} \log\frac{\mu^{2}}{\mhsq} $ 
---where $\mathcal{C}$ is the coefficient of the divergent part of the corresponding counterterm--- 
and then adding it to the amplitude.


\begin{thebibliography}{99}


\bibitem{EY} D. Espriu and B. Yencho, Phys. Rev. D 87, 055017 (2013).

\bibitem{atlas} G. Aad et al. [The ATLAS collaboration], Phys. Lett. B 716 (2012) 1. 

\bibitem{cms} S.Chatrchyan et al. [The CMS collaboration], Phys. Lett. B 716 (2012) 30.

\bibitem{ECHL} A. Dobado, D. Espriu and M.J. Herrero, Phys.Lett. B255 (1991) 405;
D. Espriu and M.J. Herrero, Nucl.Phys. B373 (1992) 117;
M.J. Herrero and E. Ruiz-Morales, Nucl.Phys. B418 (1994) 431; Nucl. Phys. B 437 (1995) 319; 
D. Espriu and J. Matias,  Phys.Lett. B341 (1995) 332;
A. Dobado, M.J. Herrero, J.R. Pel\'aez and E. Ruiz-Morales, Phys. Rev. D 62
(2000) 05501; R.~Foadi, M.~Jarvinen and F.~Sannino,
  %``Unitarity in Technicolor,''
  Phys.\ Rev.\ D {\bf 79}, 035010 (2009)
  [arXiv:0811.3719 [hep-ph]].
  %%CITATION = ARXIV:0811.3719;%%
 


\bibitem{composite} G.~F.~Giudice, C.~Grojean, A.~Pomarol and R.~Rattazzi,
  %``The Strongly-Interacting Light Higgs,''
  JHEP {\bf 0706}, 045 (2007)
  [hep-ph/0703164];  R.~Contino, M.~Ghezzi, C.~Grojean, M.~Muhlleitner and M.~Spira,
  %``Effective Lagrangian for a light Higgs-like scalar,''
  arXiv:1303.3876 [hep-ph]; R.~Alonso, M.~B.~Gavela, L.~Merlo, S.~Rigolin and J.~Yepes,
  %``The Effective Chiral Lagrangian for a Light Dynamical 'Higgs',''
  Phys.\ Lett.\ B {\bf 722}, 330 (2013)
  [arXiv:1212.3305 [hep-ph]].
  %%CITATION = ARXIV:1212.3305;%%
  %7 citations counted in INSPIRE as of 21 Jun 2013

\bibitem{bounds} 
 G.~Belanger, B.~Dumont, U.~Ellwanger, J.~F.~Gunion and S.~Kraml,
  %``Global fit to Higgs signal strengths and couplings and implications for extended Higgs sectors,''
  arXiv:1306.2941 [hep-ph];
 T.~Corbett, O.~J.~P.~Eboli, J.~Gonzalez-Fraile and M.~C.~Gonzalez-Garcia,
  %``Constraining anomalous Higgs interactions,''
  Phys.\ Rev.\ D {\bf 86}, 075013 (2012)
  [arXiv:1207.1344 [hep-ph]]; arXiv:1306.0006 [hep-ph];
 J.~Ellis and T.~You,
  %``Updated Global Analysis of Higgs Couplings,''
  arXiv:1303.3879 [hep-ph];  P.~P.~Giardino, K.~Kannike, I.~Masina, M.~Raidal and A.~Strumia,
  %``The universal Higgs fit,''
  arXiv:1303.3570 [hep-ph]; A.~Falkowski, F.~Riva and A.~Urbano,
  %``Higgs At Last,''
  arXiv:1303.1812 [hep-ph];  T.~Alanne, S.~Di Chiara and K.~Tuominen,
  %``LHC Data and Aspects of New Physics,''
  arXiv:1303.3615 [hep-ph].
  %%CITATION = ARXIV:1303.3615;%%
  %13 citations counted in INSPIRE as of 05 Aug 2013

\bibitem{nohiggs} O. Cheyette and M.K. Gaillard, Phys. Lett. B 197 (1987) 205;
A. Dobado, M.J. Herrero and T.N. Truong Phys.Lett. B235 (1990) 129; 
W. Dicus and W.W. Repko, Phys. Rev. D 42 (1990) 3660; Phys. Rev. D 44 (1991) 3473; Phys. Rev. D 47 (1993) 415;
J.R. Pel\'aez, Phys. Rev. D 55 (1997) 4193.

\bibitem{ET} J.M. Cornwall, D.N. Levin and G.Tiktopoulos, Phys. Rev. D10 (1974) 1145 ; B.W.
Lee, C. Quigg and H. Thacker, Phys. Rev. D16 (1977) 1519; G.J. Gounaris, R.
Kogerler and H. Neufeld, Phys. Rev. D34 (1986) 3257; M.S. Chanowitz and M.K. Gaillard, Nucl. Phys. B261 (1985) 379;
A. Dobado and J.R. Pel\'aez, Nucl.Phys. B425 (1994) 110; Phys. Lett. B329 (1994) 469; 
C. Grosse-Knetter and I.Kuss, Z.Phys. C66 (1995) 95;
H.J.He,Y.P.Kuang and X.Li, Phys. Lett. B329 (1994) 278.

\bibitem{esma} D. Espriu and J. Matias, Phys. Rev. D 52 (1995) 6530. 

\bibitem{iam} T.N. Truong, Phys. Rev. Lett. 61 (1988) 2526; 
A. Dobado, M.J. Herrero and T.N. Truong, Phys.Lett. B235 (1990) 134; A. Dobado and 
J.R. Pel\'aez, Phys. Rev. D 47 (1993) 4883; Phys. Rev. D 56 (1997) 3057; J.A. Oller, E. Oset 
and J.R. Pel\'aez, Phys. Rev. Lett. 80 (1998) 3452; Phys. Rev. D 59 (1999) 0740001; 60 (1999) 099906; F. Guerrero and
J.A. Oller, Nucl. Phys. B 537 (1999) 459; A. Dobado and J.R. Pel\'aez, Phys. Rev D 65 (2002) 077502.

\bibitem{DH} A. Denner, S. Dittmaier and T. Hahn, Phys.Rev. D56 (1997) 117; 
 A. Denner and T. Hahn, Nucl.Phys. B525 (1998) 27.

\bibitem{DW} S. Dawson and S. Willenbrock, Phys. Rev. D 40 (1989) 2880.

\bibitem{gupta} S.~N.~Gupta, J.~M.~Johnson and W.~W.~Repko, Phys.\ Rev.\ D {\bf 48}, 2083 (1993).

\bibitem{TGC} T.~Corbett, O.~J.~P.~Eboli, J.~Gonzalez-Fraile and M.~C.~Gonzalez-Garcia,
  %``Determining Triple Gauge Boson Couplings from Higgs Data,''
  arXiv:1304.1151 [hep-ph];
  %%CITATION = ARXIV:1304.1151;%%
  %2 citations counted in INSPIRE as of 21 Jun 2013; 
 G.~Aad {\it et al.}  [ATLAS Collaboration],
  %``Measurement of $W^+W^-$ production in $pp$ collisions at $\sqrt{s}=7$ TeV with the ATLAS detector and limits on anomalous $WWZ$ and $WW\gamma$ couplings,''
  Phys.\ Rev.\ D {\bf 87}, 112001 (2013)
  [arXiv:1210.2979 [hep-ex]];
  %%CITATION = ARXIV:1210.2979;%%
  %21 citations counted in INSPIRE as of 21 Jun 2013
 %``Measurement of $WZ$ production in proton-proton collisions at $\sqrt{s}=7$ TeV with the ATLAS detector,''
  Eur.\ Phys.\ J.\ C {\bf 72}, 2173 (2012)
  [arXiv:1208.1390 [hep-ex]];
  %%CITATION = ARXIV:1208.1390;%%
  %13 citations counted in INSPIRE as of 21 Jun 2013
 G.~Aad {\it et al.}  [ATLAS Collaboration],
  %``Measurements of Wgamma and Zgamma production in pp collisions at sqrt{s}= 7 TeV with the ATLAS detector at the LHC,''
  Phys.\ Rev.\ D {\bf 87}, 112003 (2013)
  [arXiv:1302.1283 [hep-ex]];
  %%CITATION = ARXIV:1302.1283;%%
  %6 citations counted in INSPIRE as of 21 Jun 2013
 S.~Chatrchyan {\it et al.}  [CMS Collaboration],
  %``Measurement of the sum of $W W$ and $WZ$ production with $W+$dijet events in $pp$ collisions at $\sqrt{s}=7$ TeV,''
  Eur.\ Phys.\ J.\ C {\bf 73}, 2283 (2013)
  [arXiv:1210.7544 [hep-ex]].
  %%CITATION = ARXIV:1210.7544;%%
  %9 citations counted in INSPIRE as of 21 Jun 2013





\end{thebibliography}
\end{document}